\documentclass[twocolumn]{aastex631} %

\newcommand{\kms}{{\rm km~s^{-1}}}

\newcommand{\simgt}{\,\rlap{\lower 3.5 pt \hbox{$\mathchar \sim$}} \raise
1pt \hbox {$>$}\,}
\newcommand{\simlt}{\,\rlap{\lower 3.5 pt \hbox{$\mathchar \sim$}} \raise
1pt \hbox {$<$}\,}

\newcommand{\oii}{[\textrm{O}~\textsc{ii}]}
\newcommand{\oiii}{[\textrm{O}~\textsc{III}]}

\newcommand{\hb}{${\rm H\beta}$}

\newcommand{\hst}{{HST}}

\newcommand{\Z}{{\rm [Z/H]}}

\newcommand{\gsf}{{\tt gsf}}
\newcommand{\targ}{SMACS0723}

\newcommand{\bp}{{\tt borgpipe}}
\newcommand{\sext}{{\tt SExtractor}}
\newcommand{\eazy}{{\tt EaZY}}

\newcommand{\ida}{ZD1}
\newcommand{\idb}{ZD2}
\newcommand{\idc}{ZD3}
\newcommand{\idd}{ZD4}
\newcommand{\ide}{HD1}
\newcommand{\idf}{HD2}
\newcommand{\idg}{HD3p}
\newcommand{\idh}{ZD5p}
\newcommand{\idi}{ZD6p}
\newcommand{\idj}{ZD7p}
\newcommand{\idk}{ZD8p}

\submitjournal{ApJL}

\shorttitle{Characterization of early galaxies in Webb's first deep field}
\shortauthors{Morishita \& Stiavelli}
\graphicspath{{./}{figures/}}
\begin{document}

\title{ 
Physical Characterization of { Early} Galaxies in the Webb's First Deep Field SMACS~J0723.3$-$7323
}

\author[0000-0002-8512-1404]{T. Morishita}
\affiliation{IPAC, California Institute of Technology, MC 314-6, 1200 E. California Boulevard, Pasadena, CA 91125, USA}

\author[0000-0001-9935-6047]{M. Stiavelli}
\affiliation{Space Telescope Science Institute, 3700 San Martin Drive, Baltimore, MD 21218, USA}



\begin{abstract}
This paper highlights initial photometric analyses of JWST NIRCam imaging data in the sightline of SMACS0723, aiming to identify galaxies at redshift $z>7$. By applying a conservative Lyman-break selection followed by photometric redshift analysis and visual inspection, we identify four F090W-dropout and two F150W-dropout sources, three of which were recently confirmed in an independent spectroscopic analysis to $z=7.663$, 7.665, and 8.499. We then supplement our sample with a photometric-redshift selection, and identify five additional candidates at $7<z_{\rm phot}<13$. The NIRCam images clearly resolve all sources and reveal their sub-galactic components that were not resolved/detected in the previous imaging by Hubble Space Telescope. Our spectral energy distribution analysis reveals that the selected galaxies are characterized by young stellar populations (median age of $\sim50$\,Myr) of sub-solar metallicity ($\sim0.2\,Z_\odot$) and little dust attenuation ($A_{V}\sim0.5$). In several cases, we observe extreme H$\beta$+[O~III] lines being captured in the F444W band and seen as color excess, which is consistent with their observed high star formation rate surface density. Eight of the 11 sources identified in this study appear in at least one of the recent studies of the same fields \citep{adams22,atek22,donnan22,harikane22b,yan22}, implying the high fidelity of our selection. We cross-match all high-$z$ galaxy candidates presented in the five studies with our catalog and discuss the possible causes of discrepancy in the final lists.
\end{abstract}

%Classical Novae (251) --- Ultraviolet astronomy(1736) --- History of astronomy(1868) --- Interdisciplinary astronomy(804)}


\section{Introduction} \label{sec:intro}
Our exploration of galaxies in the early universe has been enabled by the Hubble Space Telescope (\hst). Since the installation of Wide Field Camera 3 (WFC3), our redshift limit has been pushed toward $z\simgt6$, the era known as the epoch of reionization \citep[][]{gunn65,madau99,robertson15}. A tremendous amount of effort has been invested in the photometric search of early galaxies via Lyman break technique \citep{steidel96}, revealing hundreds of galaxy candidates from various HST surveys \citep[e.g.,][]{bradley12,mcleod15,bouwens15,oesch18,bowler20}. Spectroscopic followups then have successfully confirmed $\sim20$ of those candidates at $z>7$ \citep[e.g.,][]{vanzella11,shibuya12,finkelstein13,stark15,oesch15,hashimoto18,roberts-borsani22b} up to $\sim12$ \citep{oesch16,jiang21}. 

The James Webb Space Telescope (JWST) has started revolutionizing our understanding of galaxies and stellar populations in the era. With its exquisite sensitivity coverage up to $\sim5\,\mu$m, Near-Infrared Camera (NIRCam) plays the key role in pushing the redshift front of galaxy search, which enables the identification of sources beyond the previous limit \citep[][]{naidu22,castellano22,finkelstein22} up to $z\simgt20$. Identification of sources at such high redshifts is crucial to our understanding of the formation of the first galaxies and stars \citep[][]{stiavelli10,harikane22,pacucci22}.

%%%%%%%%%%%%%%%%%%%%%
\begin{figure*}
\centering
	\includegraphics[width=0.48\textwidth]{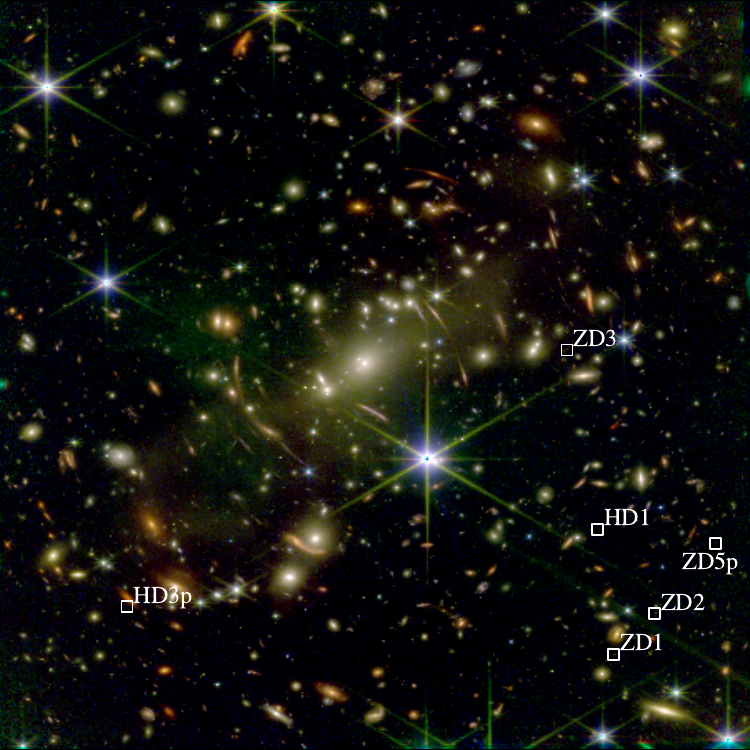}
	\includegraphics[width=0.48\textwidth]{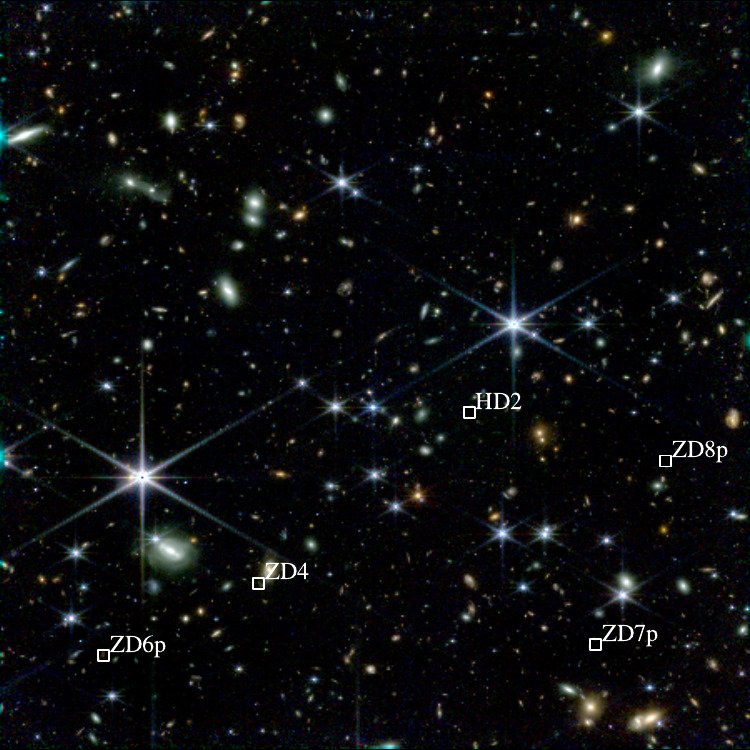}
	\caption{
    (Left): Composite RGB image of the Webb's First Deep Field, in the sightline of \targ\ (cluster field). The JWST NIRCam and NIRISS, and HST ACS and WFC3-IR images are combined (blue: F435W+F606W+F814W+F090W+F105W+F115W, green: F125W+F140W+F150W+F160W+F200W, red: F277W+F356W+F444W). The locations of the final high-$z$ galaxy candidates (Sec.~\ref{sec:view}) are marked. 
    (Right): Same but for the parallel field (blue: F090W+F150W, green: F200W+F277W, red: F356W+F444W).}
\label{fig:mosaic}
\end{figure*}
%%%%%%%%%%%%%%%%%%%%%

As part of the Early Release Observations \citep[ERO;][]{pontoppidan22}, JWST pointed to a field of \targ, a massive galaxy cluster at $z=0.390$, dubbed as the Webb's First Deep Field. This observing program, by utilizing all four instruments onboard, provides a new glimpse of the universe. The new imaging data revealed a number of potentially interesting high-redshift sources both in the cluster center and parallel fields (Fig.~\ref{fig:mosaic}). Here we present our initial identification of galaxy candidates identified by Lyman-break dropout selection at $z>7$. The simultaneous imaging of the NIRCam's dual modules allows us to explore early galaxies to: {\it i)} search for intrinsically faint-but-strongly magnified sources by gravitational lensing, and {\it ii)} provides a reference to the cluster field, as well as a glimpse of galaxy search with JWST in normal fields. This paper is structured as follows: in Sec.~\ref{sec:data}, we present our analyses on the ERO data, including initial reduction and additional photometric flux calibration. We present our selection of high-$z$ sources in Sec.~\ref{sec:sel} and analyses of their physical properties in Sec.~\ref{sec:res}. In Sec.~\ref{sec:disc}, we estimate the number density of the final candidates and discuss the fidelity of those sources by comparing identified high-$z$ sources in the same filed by other studies. Lastly, we discuss future prospects and provide the summary of this study in Sec.~\ref{sec:summary}. Throughout, we adopt the AB magnitude system \citep{oke83,fukugita96}, cosmological parameters of $\Omega_m=0.3$, $\Omega_\Lambda=0.7$, $H_0=70\,\kms\, {\rm Mpc}^{-1}$, and the \citet{chabrier03} initial mass function.

%%%%%%%%%%%%%%%%%%%%%
\begin{deluxetable}{lcc}
\tablecaption{
$5\,\sigma$-limiting magnitudes.
}
\tablehead{\colhead{ID} & \colhead{WDF-C} & \colhead{WDF-P}}
\startdata
ACS F435W & 27.2 & -- \\
ACS F606W & 28.1 & -- \\
ACS F814W & 27.5 & -- \\
WFC3 F105W & 28.1 & -- \\
WFC3 F125W & 27.9 & -- \\
WFC3 F140W & 27.9 & -- \\
WFC3 F160W & 27.8 & -- \\
NIRCam F090W & 28.4 & 28.5 \\
NIRCam F150W & 28.5 & 28.8 \\
NIRCam F200W & 28.6 & 28.9 \\
NIRCam F277W & 29.0 & 29.3 \\
NIRCam F356W & 29.2 & 29.4 \\
NIRCam F444W & 28.9 & 29.0 \\
NIRISS F115W & 28.0 & -- \\
NIRISS F200W & 27.9 & --
\enddata
\tablecomments{
Limiting magnitudes measured in $r=0.\!''16$ apertures. WDF-C and WDF-P represent the cluster and parallel fields, respectively.
}\label{tab:limmag}
\end{deluxetable}

%%%%%%%%%%%%%%%%%%%%%
\begin{figure*}
\centering
	\includegraphics[width=0.45\textwidth]{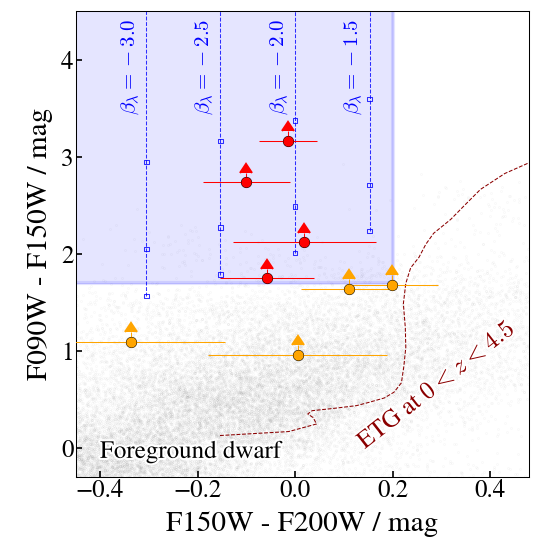}
	\includegraphics[width=0.45\textwidth]{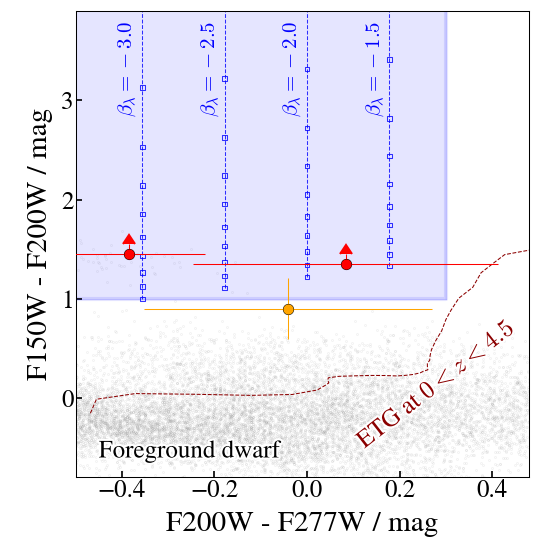}
	\caption{
	Final color-cut sample (red circles) and phot-$z$ sample (orange circles), shown in two color-color diagrams for F090W-dropout sources (left) and F150W-dropout sources (right). Color tracks of sources at various redshifts are also shown --- young galaxy template of different UV $\beta_{\lambda}$ slopes { (blue dashed lines; square symbols mark redshift, starting from $z=7.0$ and 11.8 for the left and right panels, respectively, at an increment of $\Delta z=0.1$)}, low-$z$ early-type galaxy template (red dashed lines), and foreground dwarfs (gray dots).
	}
\label{fig:cs}
\end{figure*}

%%%%%%%%%%%%%%%%%%%%%
\section{Data and Analyses}\label{sec:data}

\subsection{JWST Early Release Observations Program}\label{sec:data-jwst}
We focus our primary analysis on the NIRCam data taken as part of the JWST Early Release Observations Program, in the sightline of \targ. The NIRCam imaging data consist of F090/150/200/277/356/444W filters, with $\sim2$\,hr exposure on each. { We also include Near Infrared Imager and Slitless Spectrograph (NIRISS) imaging data (F115W and F200W), taken as part of the wide slitless spectroscopic observation in the cluster field.}

We retrieve raw data (\_uncal.fits) from a dedicated AWS storage\footnote{\url{https://outerspace.stsci.edu/display/MASTDATA/JWST+AWS+Bulk+Download+Scripts}} placed by the MAST team at STScI. We reduce raw images by using the official pipeline { (ver.1.7.2 along with the reference file context of jwst\_0988.pmap)} for detector calibration (DETECTOR) and photometric calibration (IMAGE STEP2), and then combine those in a common pixel grid (IMAGE STEP3). 

{ We made a few changes to the default pipeline processes. First, we replaced the flat-field reference files for NIRCam images with those processed and published by \citet{brammer22}.\footnote{\url{https://s3.amazonaws.com/grizli-v2/NIRCAMSkyflats/flats.html}} We find that those new flats, created with in-flight data, improve removing artifacts in the reduced images. We also adopt the magnitude zeropoints for NIRCam published by the same author.\footnote{\url{https://github.com/gbrammer/grizli/pull/107}} While the latest version of the official reference files includes a similarly improved set of magnitude zeropoints \citep{boyer22}, our choice of adopting those by \citet{brammer22} was made for the consistency with the flat files adopted above, though the difference is small ($\sim$ a few percent) across the filters between the two studies.}
Lastly, during STEP2 and STEP3, we include an extra step to subtract $1/f$-noise in \_cal.fits images by using {\tt bbpn}\footnote{\url{https://github.com/mtakahiro/bbpn}}, which follows the procedure presented in \citet{schlawin21}. The final pixel scale is set to $0.\!''0315$.

Images are aligned to the common astrometric WCS in multiple steps. First, we align the single F444W image to the GAIA-DR3 WCS frame. The rest of the NIRCam images and the NIRISS images are then aligned to the F444W image, by using both point sources and compact galaxies. We refine the pixel grid of all images to the one of F444W {\tt reproject}\footnote{\url{https://reproject.readthedocs.io/en/stable/}}. 

To homogenize the PSF of the NIRCam and NIRISS images to the one of F444W, we generate convolution kernels by following the same procedure in \citet{morishita21}. Bright stars are identified by cross-matching our sources with the GAIR DR3 catalog. We visually inspect each of them and exclude those saturated or contaminated by neighboring objects. We then resample each PSF to align those at a sub-pixel level and stacked stars of each filter to generate a median PSF. We then provide the median-stacked stars to {\tt pypher} \citep{boucaud16} and generate convolution kernels for each filter. The quality of convolved PSFs is excellent, with $\ll 1\,\%$ agreement in the encircled flux at radius $r=0.\!''16$, within which our photometric fluxes are extracted.

We also include the HST data available in the cluster field, originally taken in the RELICS program \citep{coe19,salmon20}, to supplement our selection. We retrieve raw data from MAST, which consist of seven filters (F435/606/814/105/125/140/160W) and reduce those by using \bp\ \citep{morishita21}. It is noted that the HST coverage is only for the cluster field. For the HST images, we repeat a similar analysis as for NIRCam, but convolve images to the WFC3-IR F160W PSF, which has a similar FWHM as of F444W. This choice was made because of significantly different PSF shapes in JWST and HST images. While this may leave systematic offsets between the two instruments, those will be removed in the following analysis (Sec.\ref{sec:zp}). 

For photometric flux extraction, we run \bp, a photometric pipeline designed for HST and JWST data reduction and photometric analyses. \bp\ runs \sext\ \citep[ver.2.25.0;][]{bertin96} for image detection and flux extraction, while it includes extra steps for sophisticated error estimate of images, aperture correction, and Lyman-break dropout selection of high-$z$ candidates, as done in similar previous studies \citep{trenti12,morishita18b}. We create infrared stacks of the F277/356/444W filters as the detection image for both fields. { We set configuration parameters of \sext\ as follows: DETECT\_MINAREA 0.0162\,arcsec$^2$, DETECT\_THRESH 1.5, DEBLEND\_NTHRESH 64, DEBLEND\_MINCONT 64, BACK\_SIZE 20, and BACK\_FILTSIZE 3.} The photometric magnitude limits measured with $r=0.\!''16$ apertures are listed in Table~\ref{tab:limmag}.

%%%%%%%%%%%%%%%%%%%%%
\subsection{Zero-point correction across filters and instruments}\label{sec:zp}
Next, we aim to calibrate flux zeropoints of imaging data, including offsets caused by the different PSF profiles of F444W and F160W. We first correct the relative flux offset between the two telescopes by scaling pseudo F150W fluxes, interpolated from the observed HST F140W and F160W fluxes, to NIRCam F150W. We use bright and isolated objects with $S/N>20$ taken from the photometric catalog constructed in the previous section. For the pseudo F150W fluxes, we assume a linear slope between the two HST filters and use the value at the wavelength center of the NIRCam F150W filter. We adopt the median value of the correcting factor, $f_{150}/f_{140+160}=1.264$. We then run {\tt eazypy}, a python wrapper of photometric redshift code \eazy\ \citep{brammer08}, to fine-tune magnitude zeropoints across all filters. We run redshift fitting on those with spectroscopic redshifts \citep{mahler22}, both cluster member galaxies and background emission line galaxies. The derived correction factor is $-1.5$--$4.4\,\%$ relative to F150W as the pivot { (Table~\ref{tab:magzpcor})}, requiring only minor correction.\footnote{The derived correction factors were $30$--$40\%$ in the NIRCam filters in our original analysis using the calibration context of {\tt jwst\_0916.pmap} and the ver.1.6.0 pipeline and consistent with those reported in \citet{rigby22}.} 
We apply the correction to the photometric catalog. For the NIRCam images in the parallel field, where no spectroscopic measurement or HST images are available, we apply the same zeropoint-offsets derived for the cluster field.

%=============================
\begin{deluxetable}{llc}%[!h]
\tabletypesize{\footnotesize}
\tablecolumns{3}
\tablewidth{0pt} 
\tablecaption{
Magnitude-zeropoint correction factors (Sec.~\ref{sec:zp}).
}
\tablehead{
\colhead{Instrument} & \colhead{Filter} & \colhead{Correction factor}\\
\colhead{} & \colhead{} & \colhead{}
}
\startdata
ACS & F435W & 1.025\\
ACS & F606W & 1.009\\
ACS & F814W & 1.032\\
WFC3 & F105W & 1.017\\
WFC3 & F125W & 1.000\\
WFC3 & F140W & 1.044\\
WFC3 & F160W & 0.994\\
NIRCam & F090W & 1.016\\
NIRCam & F150W & 1.000\\
NIRCam & F200W & 1.029\\
NIRCam & F277W & 1.016\\
NIRCam & F356W & 1.016\\
NIRCam & F444W & 1.032\\
NIRISS & F115W & 0.985\\
NIRISS & F200W & 0.985
\enddata
\tablecomments{
NIRCam F150W filter is used as the pivot point. For HST filters, the correction factors here are applied after the universal correction of $1.264$ for aperture correction (Sec.~\ref{sec:data}).
}
\label{tab:magzpcor}
\end{deluxetable}
%=============================

%%%%%%%%%%%%%%%%%%
\section{Selection of high-redshift galaxy candidates}\label{sec:sel}
In what follows, we present our selection of galaxy candidates at $z\simgt7$. { To make our identification of high-$z$ candidates as comprehensive as possible, we adopt two selection methods --- conventional color-cut selection and supplemental photometric-redshift (phot-$z$) selection. The former provides a sample of high purity, whereas the latter covers fainter sources that fall outside the color-cut boundary.}

%%%%%%%%%%%%%%%%%%
\subsection{Color-cut Selection}\label{sec:cs}
Our first selection consists of three steps --- Lyman break color-cut selection, photometric redshift analysis, and visual inspection. For the filters available in this study, we explore four different redshift ranges. 

%%%%%%%%%%%%%%%%%%
\subsubsection{Lyman-break color cut}\label{sec:lyb}
We select candidate galaxies in the following { four redshift ranges enabled by six NIRCam filters available in both fields:

$F090W-dropouts$:
$$S/N_{\rm 150}>3.0$$
$$m_{090}-m_{150}>1.7$$
$$m_{150}-m_{200}<0.2$$
$$S/N_{\rm 090 (, 435, 606, 814)}<2.0$$

$F150W-dropouts$:
$$S/N_{\rm 200}>3.0$$
$$m_{150}-m_{200}>1.0$$
$$m_{200}-m_{277}<0.3$$
$$S/N_{\rm 090 (, 435, 606, 814, 105, 115, 125)}<2.0$$

$F200W-dropouts$:
$$S/N_{\rm 277}>3.0$$
$$m_{200}-m_{277}>1.6$$
$$m_{277}-m_{356}<0.25$$
$$S/N_{\rm 090, 150 (, 435, 606, 814, 105, 115, 125, 140, 160)}<2.0$$

$F277W-dropouts$:
$$S/N_{\rm 356}>3.0$$
$$m_{277}-m_{356}>1.4$$
$$m_{356}-m_{444}<0.25$$
$$S/N_{\rm 090, 150, 200 (, 435, 606, 814, 105, 115, 125, 140, 160)}<2.0$$
}
In addition, we also impose $S/N>4$ in detection images for all selections above, to minimize the fraction of artifacts.

When NIRISS or HST images are available (i.e. the cluster field), we supplement our selection by adding those to the non-detection filters. The color boundary of each selection is designed to effectively capture young, relatively dust-free galaxies at $7\simlt z\simlt 11$, { $11\simlt z\simlt 15$, $15\simlt z\simlt 20$, and $21\simlt z\simlt 28$, respectively.} 

It is noted that in the F090W-dropout selection, we require non-detection in F090W ($z$-band dropout) while we calculate the color of Lyman break by using the flux upper limit estimated in the image in the photometric step above; thus, F090W$-$F150W color of the sources selected in this selection is an upper limit (Fig.~\ref{fig:cs}).

In Fig.~\ref{fig:cs}, we show color trajectories of three power-law UV slopes, $\beta_\lambda=-1.5,-2,-2.5$, and $-3$). Our color-cut selection is designed to effectively minimize low-$z$ interlopers, such as foreground dwarfs and low-$z$ galaxy populations. Colors of dwarf stars, taken from the IRTF spectral library \citep{rayner03} and the SpeX prism library \citep{burgasser14}, and galaxies at $0<z<4.5$ are shown. 

%%%%%%%%%%%%%%%%%%%%
\subsubsection{Photometric-redshift cut}\label{sec:ps}
We apply photometric redshift cut to those selected in the previous color-cut section, in a similar same way as performed in previous work \citep{morishita18b,roberts-borsani22,ishikawa22}. This is to minimize the fraction of low-$z$ interlopers such as galaxies with old stellar populations \citep[e.g.,][]{oesch16} and foreground dwarfs \citep[e.g.,][]{morishita20} that may migrate to the color boundary box due to photometric scatters. We run \eazy\ with the default magnitude prior set to the filter that covers the wavelength of Lyman break. The redshift range is set to $0<z<30$, with a step size of $\log (1+z) = 0.01$. By following \citet{morishita18b}, we exclude those with $p(z<6.5)>0.2$, i.e. total redshift probability at $z<6.5$ is greater than $20\,\%$.

%%%%%%%%%%%%%
\begin{deluxetable*}{rccclccll}
\tablecaption{List of our final candidates.}
\tabletypesize{\footnotesize}
\tablecolumns{9}
\tablewidth{0pt} 
\tablehead{
\colhead{ID} & \colhead{R.A.} & \colhead{Decl.} & \colhead{$m_{\rm UV}$} & \colhead{$z_{\rm phot.}$} & \colhead{$p(z>6.5)$} & \colhead{$\mu$$^\dagger$} & \colhead{Comments} & \colhead{Claimed$^\ddagger$}\\
\colhead{} & \colhead{degree} & \colhead{degree} & \colhead{mag} & \colhead{} & \colhead{} & \colhead{} & \colhead{} & \colhead{}
}
\startdata
\cutinhead{F090W-dropouts ($7\simlt z\simlt 11$)}
WDF-C-769 / \ida & 110.8343506 & -73.4345093 & 26.6 & $7.34_{-0.07}^{+0.05}$ & 1.00 & 1.54 & $z_{\rm spec.}=7.663$ & [1],[2]\\
WDF-C-1045 / \idb & 110.8449173 & -73.4350433 & 26.0 & $8.55_{-0.18}^{+0.15}$ & 1.00 & 1.56 & $z_{\rm spec.}=7.666$& [1],[2]\\
WDF-C-3186 / \idc & 110.8598175 & -73.4491272 & 27.1 & $9.03_{-0.24}^{+0.31}$ & 1.00 & 15.3 & $z_{\rm spec.}=8.498$ & [1],[2],[3]\\
WDF-P-1762 / \idd & 110.6461792 & -73.4758453 & 27.9 & $10.27_{-0.64}^{+0.71}$ & 1.00 & -- & & [1],[3]\\
\cutinhead{F150W-dropouts ($10\simlt z\simlt 14$)}
WDF-C-1730 / \ide & 110.8452759 & -73.4404297 & 28.1 & $13.85_{-1.39}^{+1.36}$ & 0.81 & 2.00 & & [4]\\
WDF-P-3504 / \idf & 110.6964722 & -73.4768295 & 28.5 & $13.41_{-1.56}^{+1.25}$ & 0.98 & -- & & \\
\cutinhead{Phot-z sample}
WDF-C-1622 / \idh & 110.8616486 & -73.4362259 & 28.2 & $8.21_{-0.79}^{+1.02}$ & 1.00 & 1.63 & & [3]\\
WDF-P-949 / \idi & 110.6147919 & -73.4774246 & 27.7 & $10.87_{-0.64}^{+0.65}$ & 1.00 & -- & Combined with WDF-P-963.& [1]\\
WDF-P-963 / \idi & 110.6146545 & -73.4773712 & 28.2 & $7.35_{-0.18}^{+3.65}$ & 0.99 & -- & Combined with WDF-P-949.& [1]\\
WDF-P-1095 / \idj & 110.6909027 & -73.4629288 & 29.0 & $10.58_{-1.35}^{+0.97}$ & 0.97 & -- & & [3]\\
WDF-P-3004 / \idk & 110.7211685 & -73.4687576 & 28.6 & $10.02_{-1.51}^{+1.01}$ & 0.84 & -- & &\\ WDF-C-1152 / \idg & 110.7653885 & -73.4514236 & 27.9 & $12.70_{-1.31}^{+2.05}$ & 0.81 & 4.72 &Close to a dusty galaxy at $z\sim2.5$& \\
\enddata
\tablecomments{
For the color cut selection, no dropout sources other than F090W-dropouts were identified. Photometric redshift uncertainty is estimated by only using the probability distribution at $z>6.5$. WDF-P-949 and WDF-P-963 are treated as a single system in the main text (Sec.~\ref{sec:pzsample}).
{ $\dagger$ Median magnification of various lens models \citep{jullo07,oguri10,caminha22,mahler22} calculated at the position of the source.}
$\ddagger$ Cross-appearance of the source in other studies. 1:\citet{adams22} 2:\citet{atek22} 3:\citet{donnan22} 4:\citet{yan22}. None of our samples appears in \citet{harikane22b}. 
}\label{tab:allcand_phot-z}
\end{deluxetable*}

%%%%%%%%%%%%%%%
\subsubsection{Visual Inspection}\label{sec:vi}
Lastly, we visually inspect the sources remaining after the previous two steps. In this step, we exclude those near diffraction spikes or possible artifacts, such as flux residuals of snowballs \citep{rigby22}. Sources are excluded when either of the inspectors (TM, MS) raises a flag that the object is an obvious artifact, blending with bright sources (where flux measurement would be challenging), or shows a significant amount of positive pixels in the non-detection filters.

%%%%%%%%%%%%%%%
{
\subsection{Supplemental phot-$z$ selection}\label{sec:addsele}
While the source selection based on the color cut is designed to provide a sample of high purity, it may not cover all possible high-$z$ sources, especially those 1.~near the color selection window or 2.~faint, whose Lyman break color is a relatively weak upper limit. To improve the completeness, we here aim to identify possible sources as a supplemental sample. 

We run \eazy\ as in Sec.~\ref{sec:zp} on those detected at $S/N>4$ in the IR-detection filter and then select those with $p(z>6.5)>0.8$. To further eliminate low-$z$ interlopers, we impose non-detection ($S/N<2$) in all filters that are blueward to the Lyman break inferred from the derived redshift. We then visually inspect individual candidates, to exclude any suspicious objects and artifacts. 
}

%%%%%%%%%%%%%%%%%%%%%
\begin{figure*}
\centering
	\includegraphics[width=.89\textwidth]{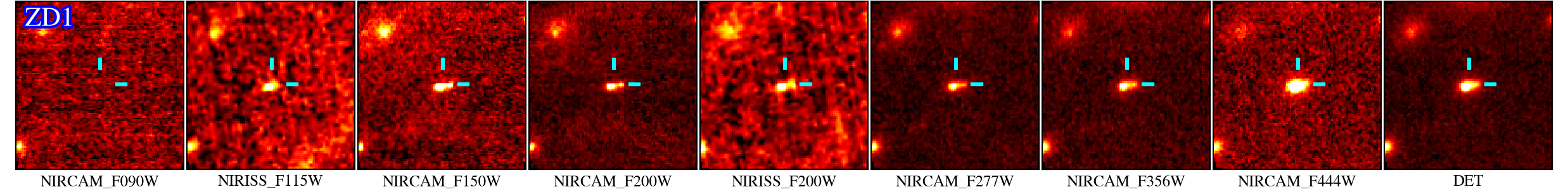}
	\includegraphics[width=.89\textwidth]{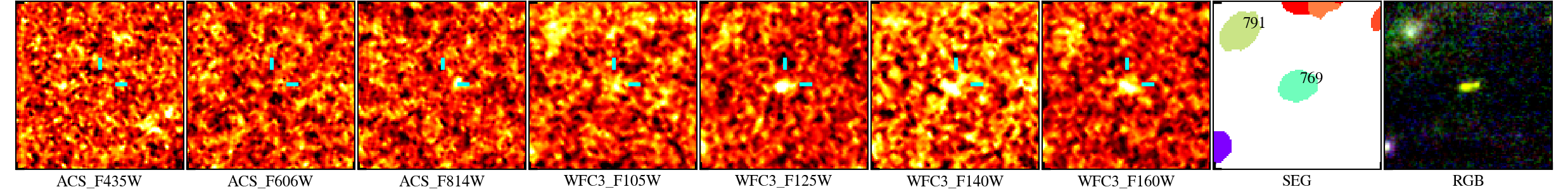}
	\includegraphics[width=.89\textwidth]{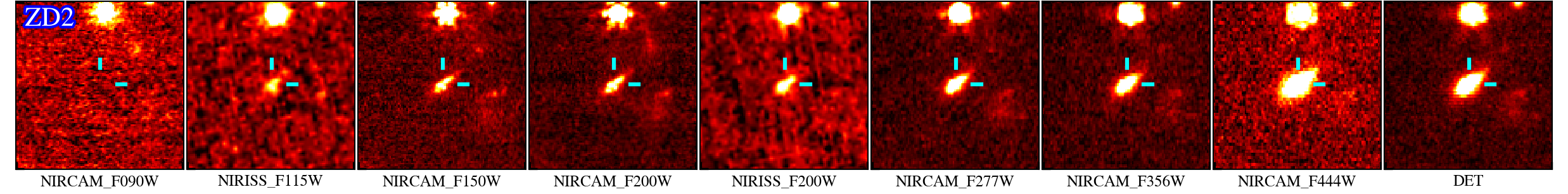}
	\includegraphics[width=.89\textwidth]{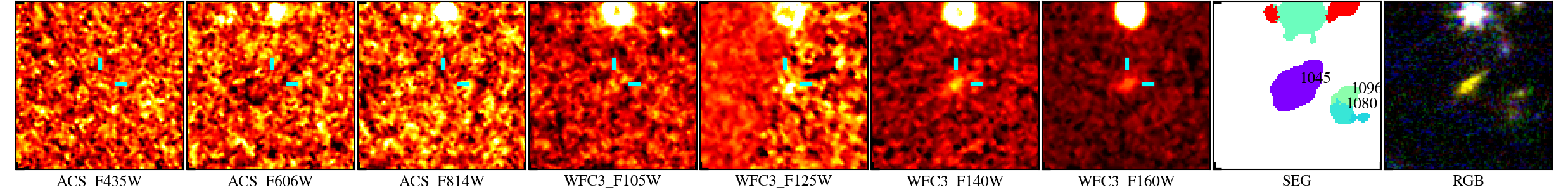}
	\includegraphics[width=.89\textwidth]{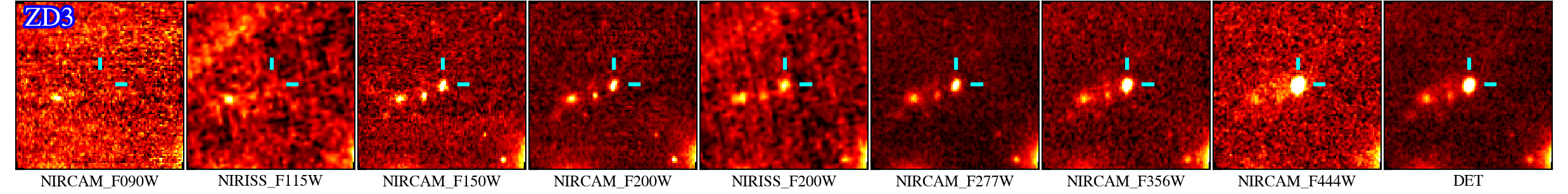}
	\includegraphics[width=.89\textwidth]{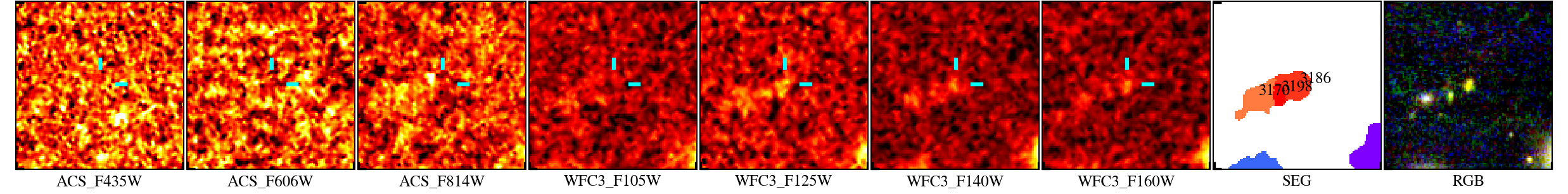}
	\includegraphics[width=.89\textwidth]{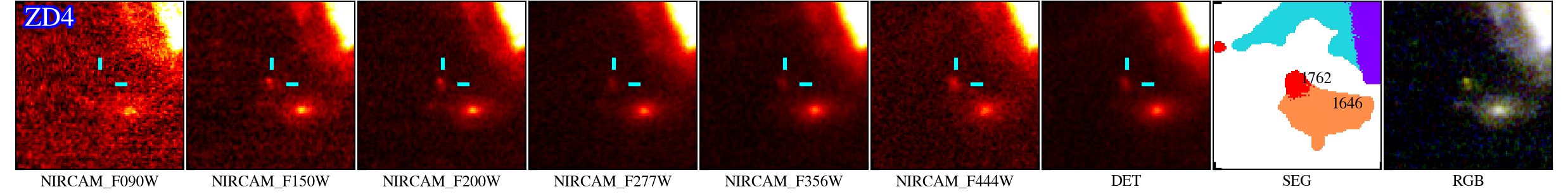}
	\includegraphics[width=.89\textwidth]{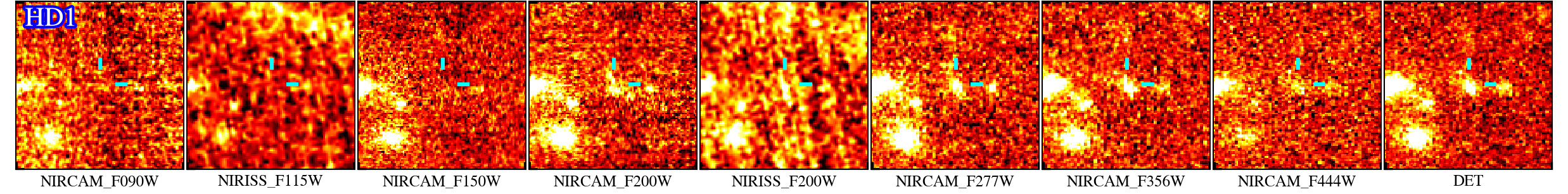}
	\includegraphics[width=.89\textwidth]{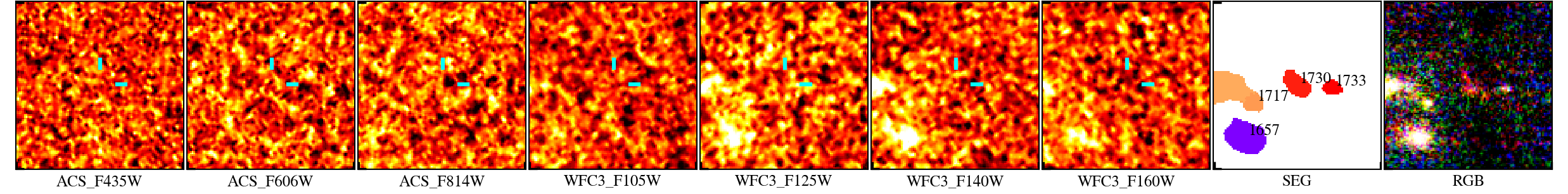}
	\includegraphics[width=.89\textwidth]{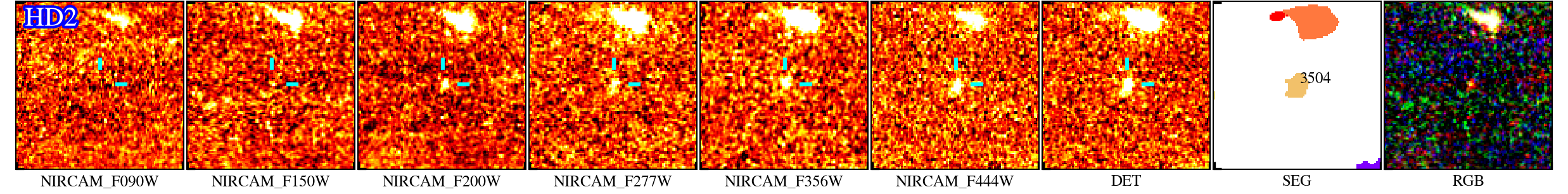}
	\caption{
	Postage stamps of the final color-cut sample (F090W-dropouts for the first four and F150W-dropouts for the following two). For each object, the top panel shows the image stamps ($3.2\times 3.2$\,arcsec$^2$). `DET' represents the IR-stacked image. The source detection segmentation map and rgb composite (NIRCam F090W/F150W/F200W) image are also shown in the two right panels. 
    }
\label{fig:mosaic-f090w-cls}
\end{figure*}

%%%%%%%%%%%%%%%%%%%%%
\begin{figure*}
\centering
	\includegraphics[width=.89\textwidth]{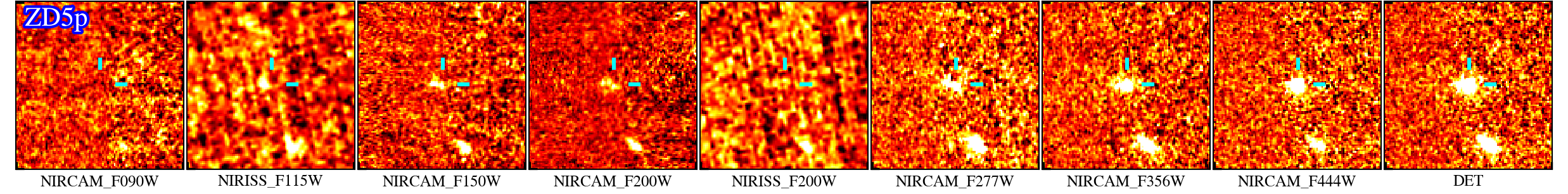}
	\includegraphics[width=.89\textwidth]{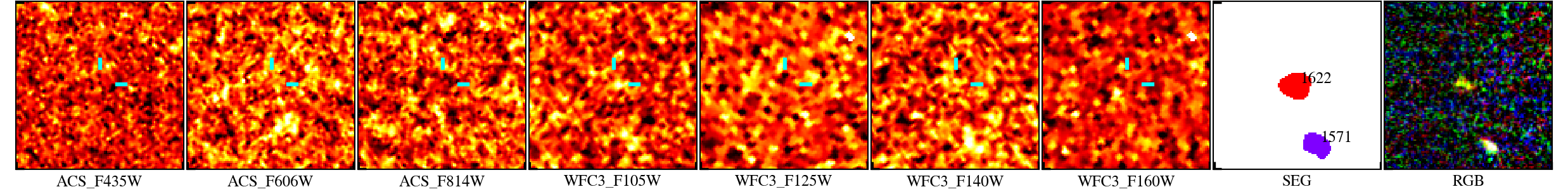}
	\includegraphics[width=0.89\textwidth]{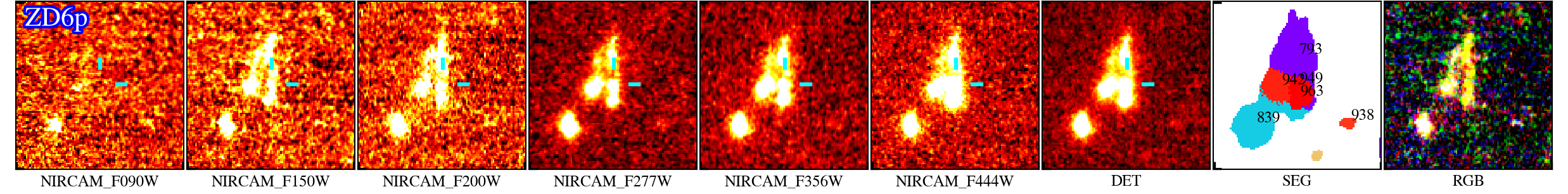}
	\includegraphics[width=0.89\textwidth]{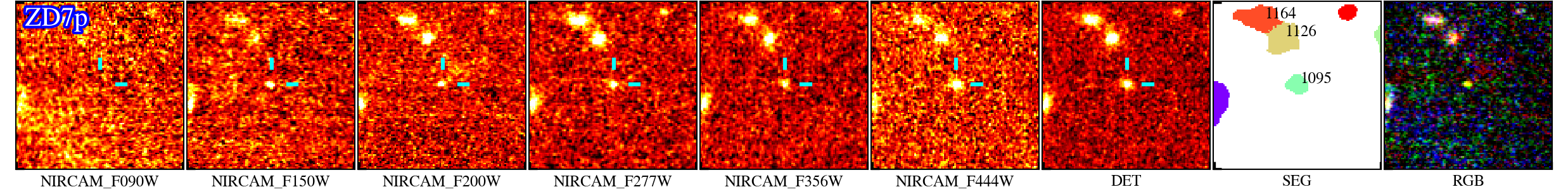}
	\includegraphics[width=0.89\textwidth]{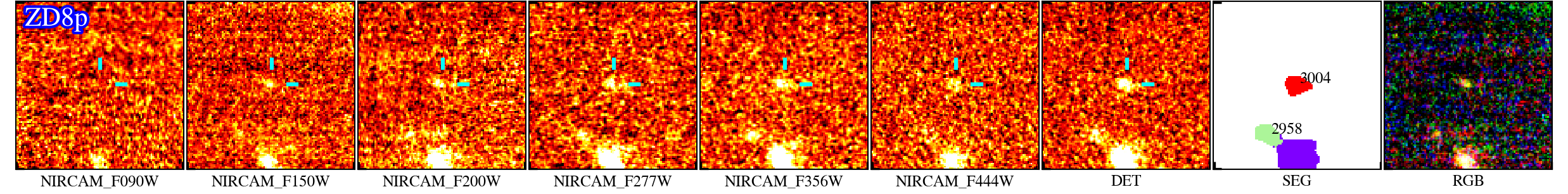}
	\includegraphics[width=.89\textwidth]{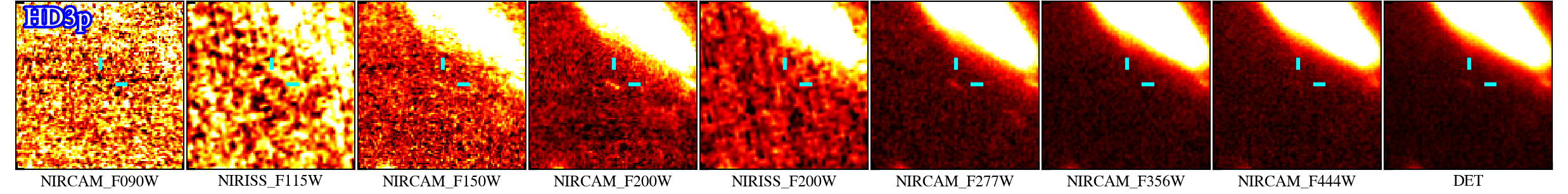}
	\includegraphics[width=.89\textwidth]{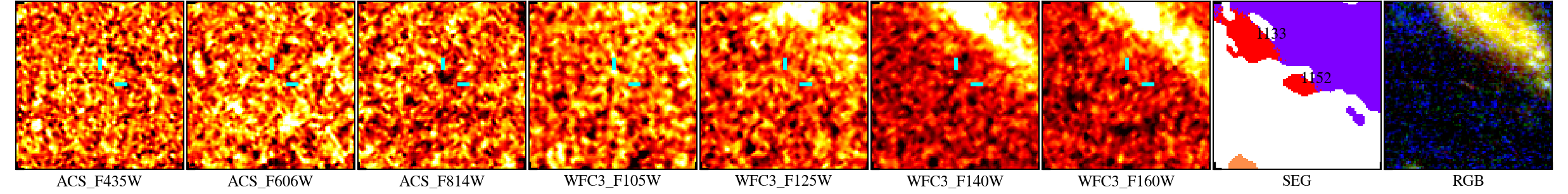}
	\caption{
	Same as Fig.~\ref{fig:mosaic-f090w-cls} but for the phot-$z$ sample. IDs~949 and 963 are analyzed and presented as a single system, \idi.
    }
\label{fig:mosaic-f090w-par}
\end{figure*}

%%%%%%%%%%%%%%%%%
\section{Results}\label{sec:res}
{ From the color-cut selection, we identified four F090W-dropouts, two F150W-dropouts, and none from the higher redshift bins (Fig.~\ref{fig:mosaic-f090w-cls}). From the phot-$z$ selection, we identify additional five sources at $7<z<11$ (Fig.~\ref{fig:mosaic-f090w-par}).} Our final candidates are summarized in Table~\ref{tab:allcand_phot-z}. 

In what follows, we provide an overview of the final candidates and their physical properties obtained by spectral energy distribution (SED) and morphological analyses. 
{ For those in the cluster field, magnification factors are estimated by taking the median value of various lens models \citep{jullo07,oguri10,caminha22,mahler22} at the position of each source.}

%%%%%%%%%%%%%%%%%
\subsection{Overview of selected candidate galaxies}\label{sec:view}

%%%%%%%%%%%%%%%%%%%%%%
\subsubsection{The color-cut sample}\label{sec:ccsample}
In Fig.~\ref{fig:mosaic-f090w-cls}, we show postage stamps of the final color-cut samples. For the five sources in the cluster field, they are not detected in the optical filters of HST in addition to their clear dropout in the F090W band. Indeed, all three of those galaxies are spectroscopically confirmed in recent work \citep{carnall22,tacchella22}, to $z=7.663$ (\ida), 7.665 (\idb), and 8.498 (\idc).

\begin{figure*}
\centering
	\includegraphics[width=0.48\textwidth]{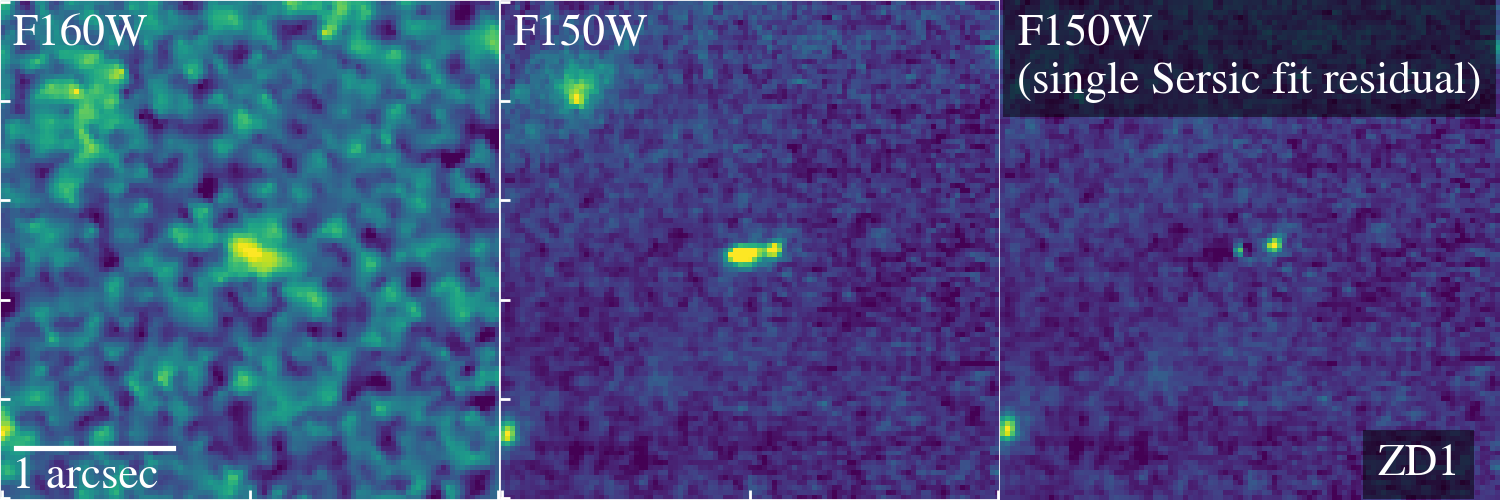}
	\includegraphics[width=0.48\textwidth]{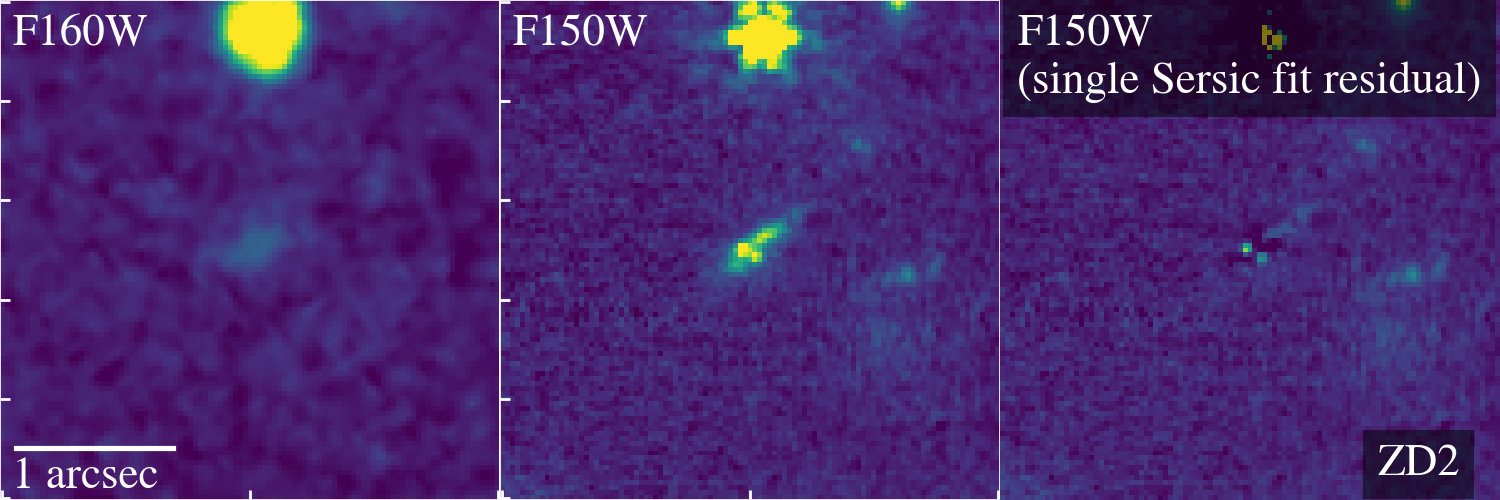}
	\includegraphics[width=0.48\textwidth]{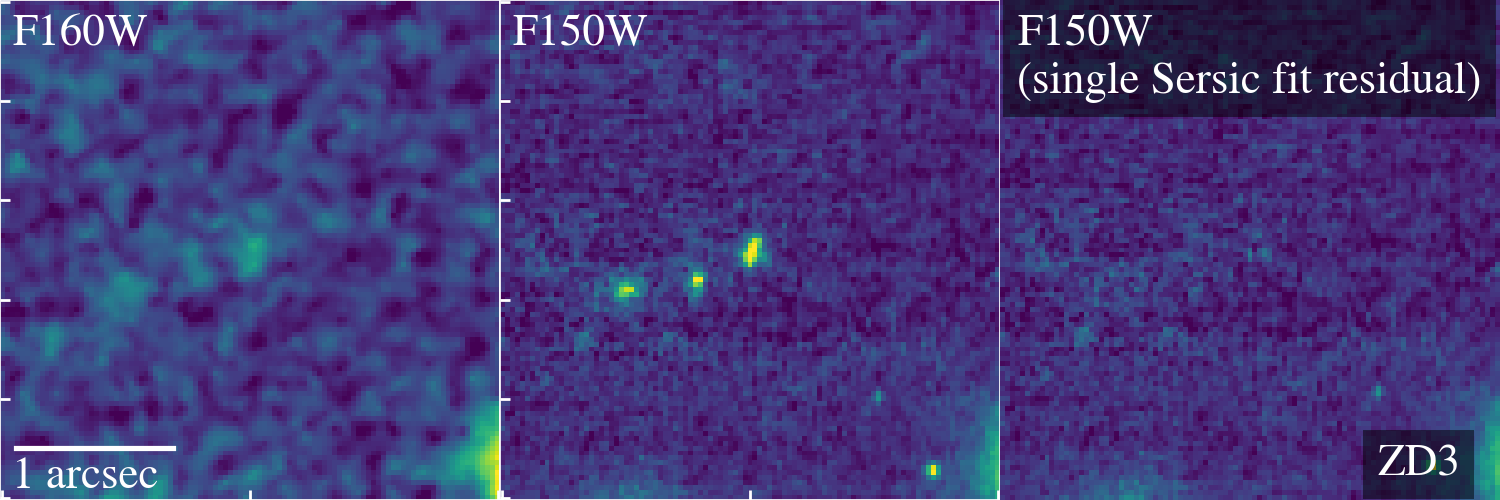}
	\includegraphics[width=0.48\textwidth]{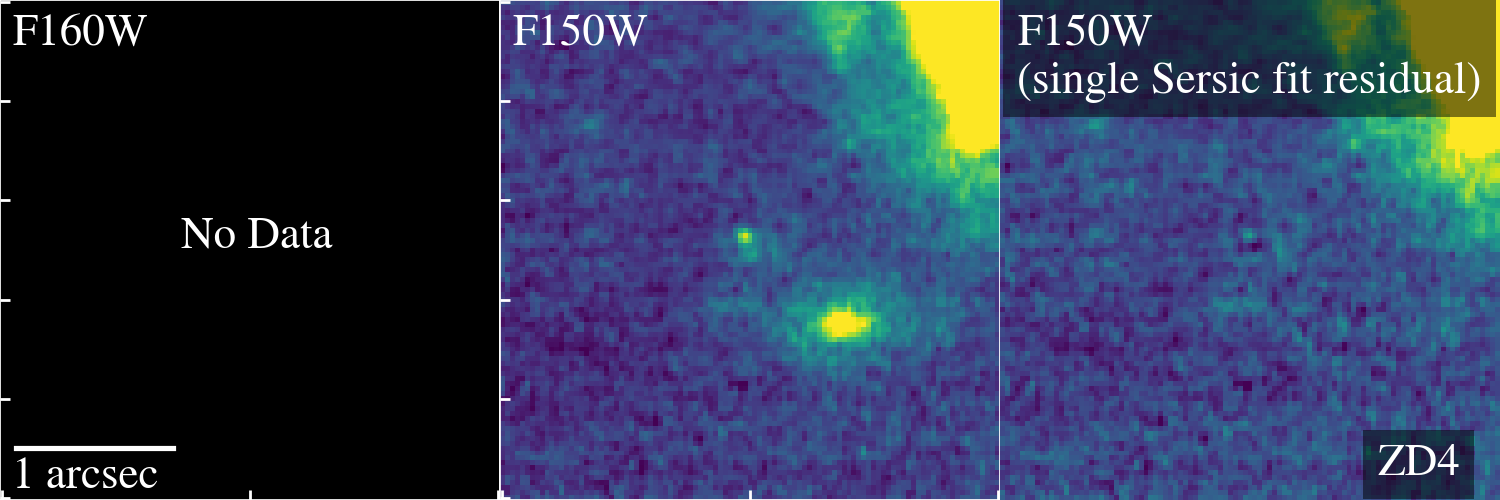}
	\includegraphics[width=0.48\textwidth]{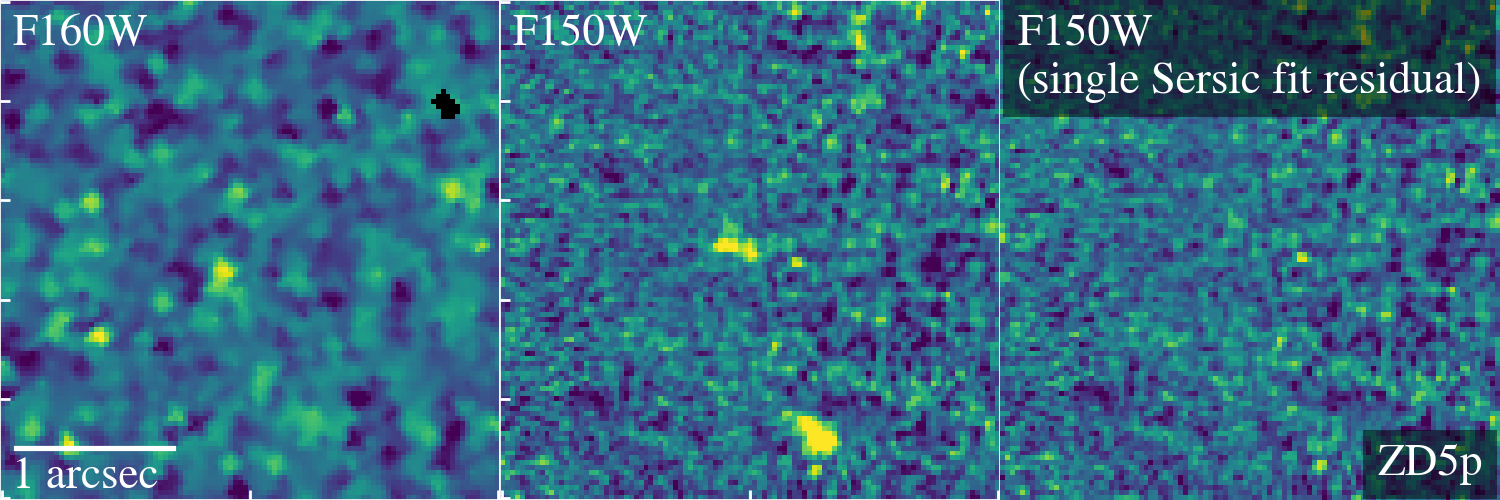}
	\includegraphics[width=0.48\textwidth]{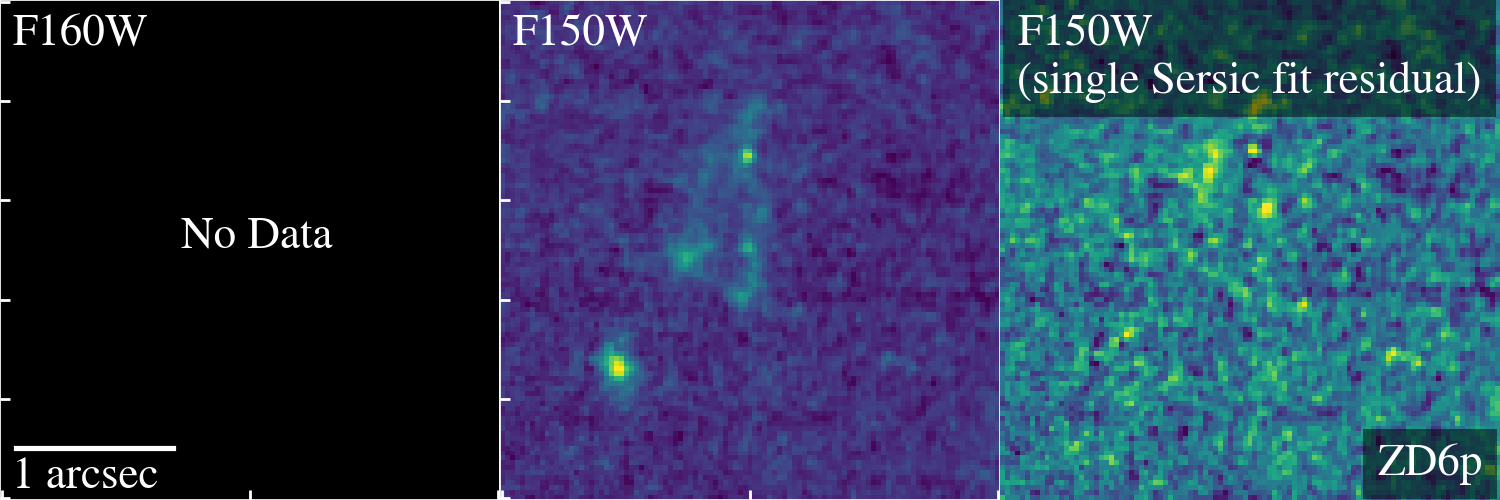}
	\includegraphics[width=0.48\textwidth]{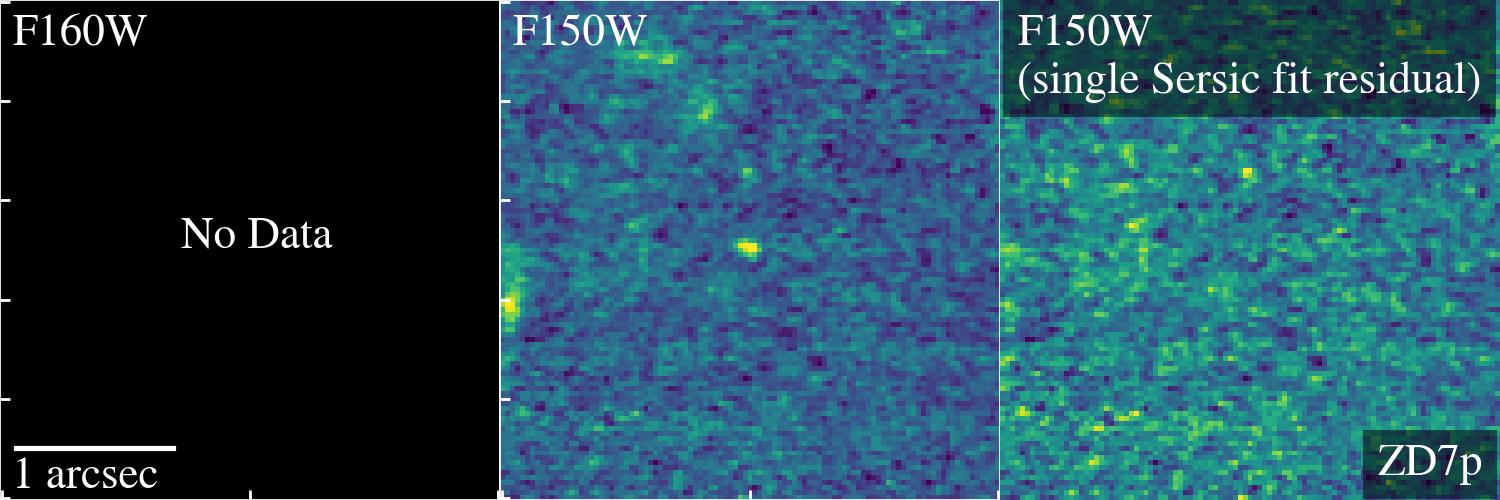}
	\includegraphics[width=0.48\textwidth]{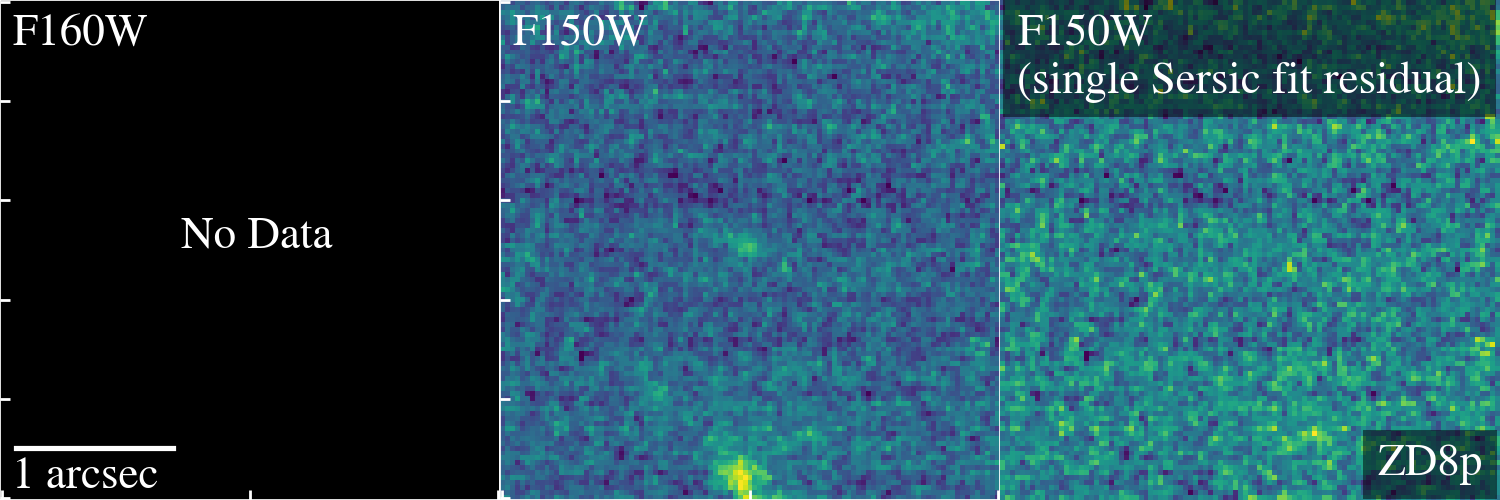}
	\includegraphics[width=0.48\textwidth]{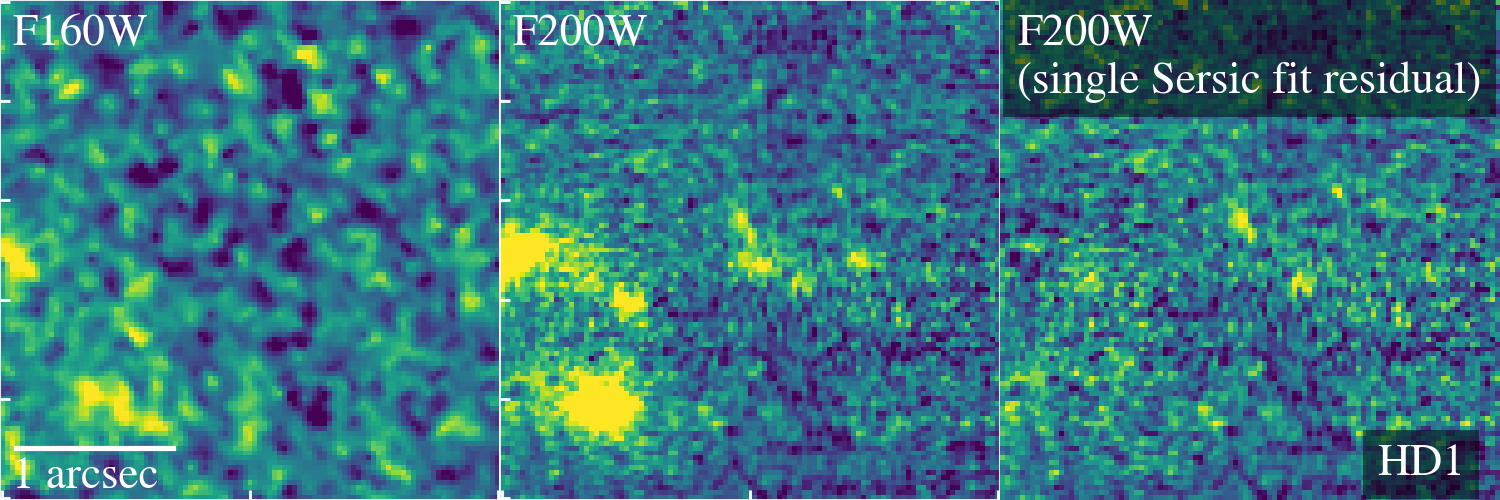}
	\includegraphics[width=0.48\textwidth]{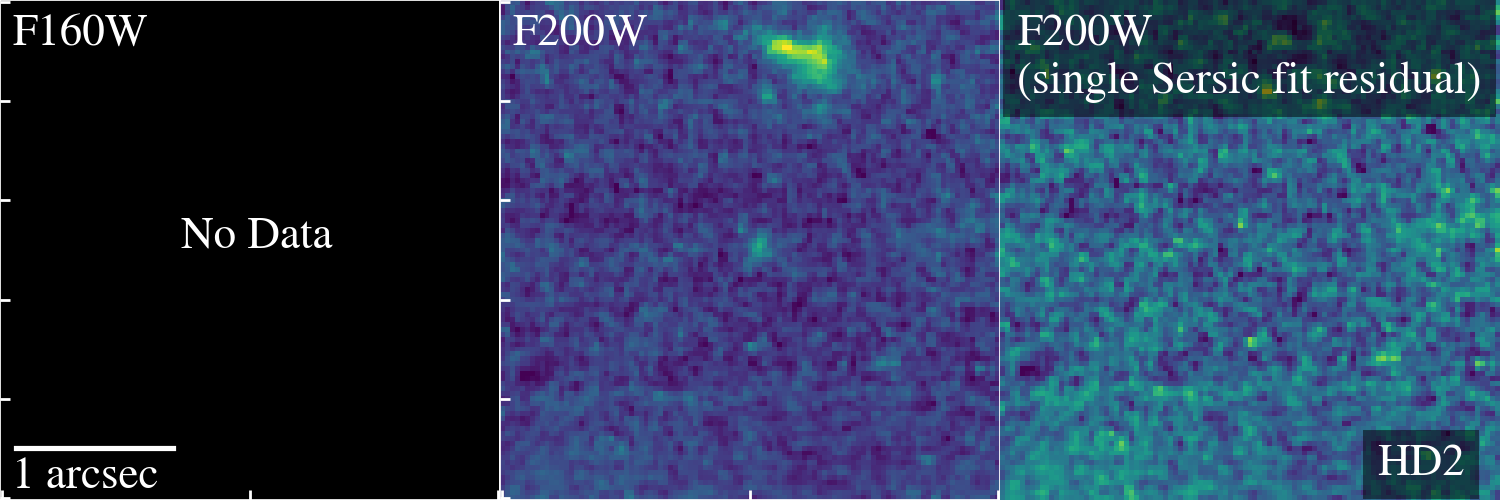}
	\includegraphics[width=0.48\textwidth]{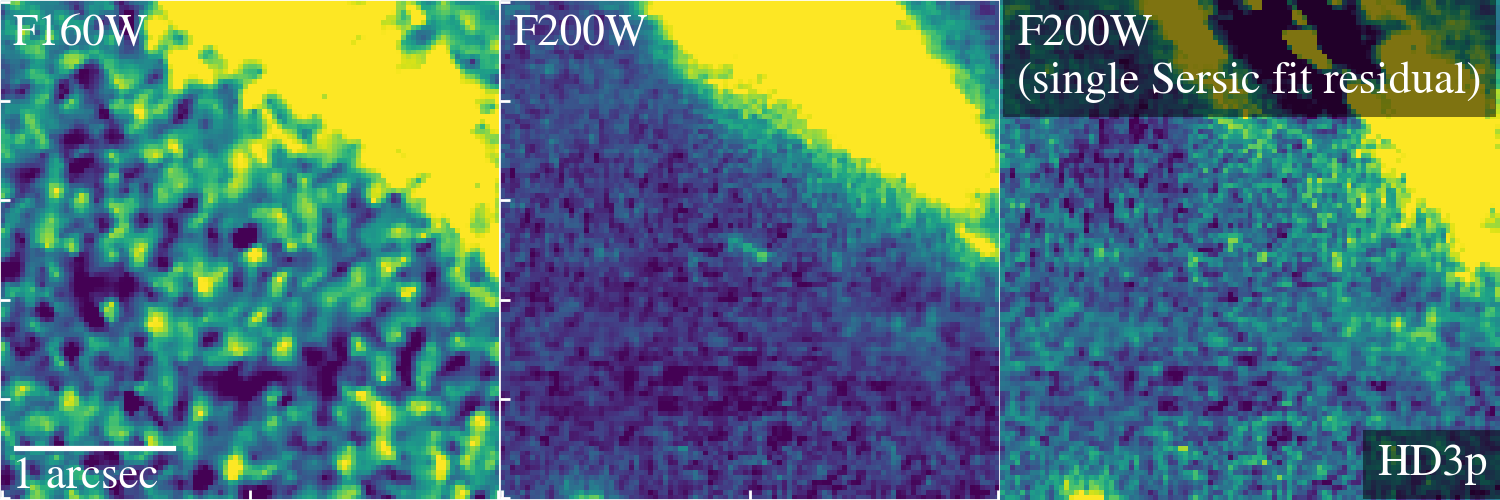}
	\caption{
    Zoom-in image ($3.\!''2\times3.\!''2$) of the final candidates at $z<12$ (in NIRCam F150W) and $z>12$ (NIRCam F200W). HST WFC3-IR F160W is shown for those in the cluster field. In the right panel we show the residual image after subtracting a single S\'ersic component in the NIRCam image. Surrounding objects are either fitted simultaneously or masked.
	}
\label{fig:gf}
\end{figure*}

The NIRCam images resolve all candidates and reveal their morphology in detail in rest-frame UV ($\sim0.15\,\mu$m in F090W) to optical ($\sim0.44\,\mu$m in F444W). The comparison of the NIRCam images to WFC3-IR images clearly showcases the resolving power of JWST. Aside from multi-band postage stamps above, in Fig.~\ref{fig:gf} we also show the zoom-in image of each candidate in WFC3 F160W and NIRCam images. 

\ida\ is isolated from other neighboring sources. The source consists of one primary core and a small component to its right. To provide a detailed view of this sub-component, we perform single component S\'ersic fit in the F150W image (Fig.~\ref{fig:gf}) using {\tt galfit} \citep{peng02}, where we clearly see non-negligible residuals. In the stamp of \idb, there are two sources to the lower-right direction but they are faint and reasonably apart from \idb, and their flux contamination to \idb\ is considered small. Same as for \ida, single S\'ersic fit reveals multiple components, including a diffuse extended component. \idc\ has two compact companions that align in one direction. \sext\ successfully deblends each of the three sources. Among the three sources, \idc\ is the brightest and thus the flux contamination is also considered small. Indeed, S\'ersic fit to \idc, with the two close companions included simultaneously, reveals smoother residuals compared to the other two. Thus, \idc\ is characterized as compact, despite its higher magnification. We present the measured sizes in Table~\ref{tab:prop}. The middle component (ID 3198 in the segmentation map) exhibits similar dropout color as for \idc. However, we find that this object is detected in F105W, whereas \idc\ is undetected in the same band. In addition, while uncertainty is large, the photometric redshift of ID~3198 is $z\sim1.8$. We thus conclude that, despite the proximity and the similarity in color, this object is not likely at the same redshift.

\idd\ is identified in the parallel field, with photometric redshift of $z_{\rm phot.}\sim10.3\pm0.7$. While the photometric redshift distribution extends a relatively wider range, the solution is unique to high-$z$, with $p(z<6.5)\ll1\%$. \idd\ shows extended morphology, elongated to the vertical direction. Interestingly, it shows two separate components in the F150W and F200W filters, while it is barely separated in the long channel of NIRCam, possibly due to lower spatial resolution. Since our detection is based on the F444W image, these (possible) two components are not deblended in our catalog. 

{ In the F150W-dropout selection, we identify two candidates. \ide\ shows clear non-detection in F090W, F115W, and F150W filters. The object is resolved and shows multiple components, while those components are not deblended in our detection configuration. \idf\ is a faint object and flux measurements come with relatively large errors. Despite, the measured F150W-F200W color satisfies the color criterion, with the redshift probability of $p(z>6.5)=0.98$. It is noted that while the F150W image shows some positive pixels at the location of the object, non-detection in the band is not required in this selection. 
}

Among the color-cut objects identified in the cluster field, \idc, \idf, and \idg\  were not detected in the public photometric catalog by the RELICS team\footnote{\url{https://relics.stsci.edu}}. 
\ida\ and \idb\ are detected and listed in the catalog (with their IDs 62 and 75) but with photometric redshift of $z\sim1.6$ and $2.0$. Thus, none of the four galaxies was identified as a candidate high-$z$ source from the HST-only dataset. { It is worth mentioning that the two objects are flagged as detected in the ACS F606W filter in the RELICS catalog, with $2.5$ and 3.1\,$\sigma$, respectively, while not in the other two ACS filters (F435W and F814W). By revisiting their original F606W image, we could not confirm any convincing detection but only a small fraction of positive pixels near the position in both cases. The false detection might have rejected those objects as a high-$z$ candidate in their original selection.}

%%%%%%%%%%%%%%%%%%%%%

%%%%%%%%%%%%%%%%%%%%%
\begin{figure*}
\centering
	\includegraphics[width=.49\textwidth]{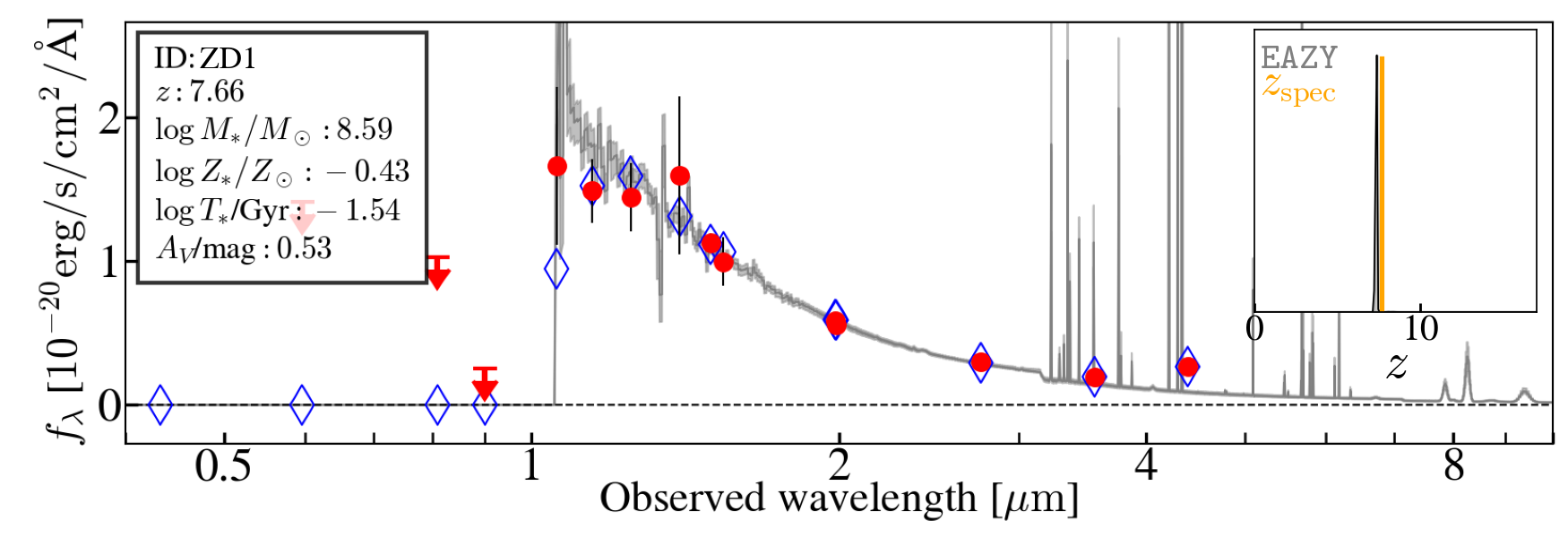}
	\includegraphics[width=.49\textwidth]{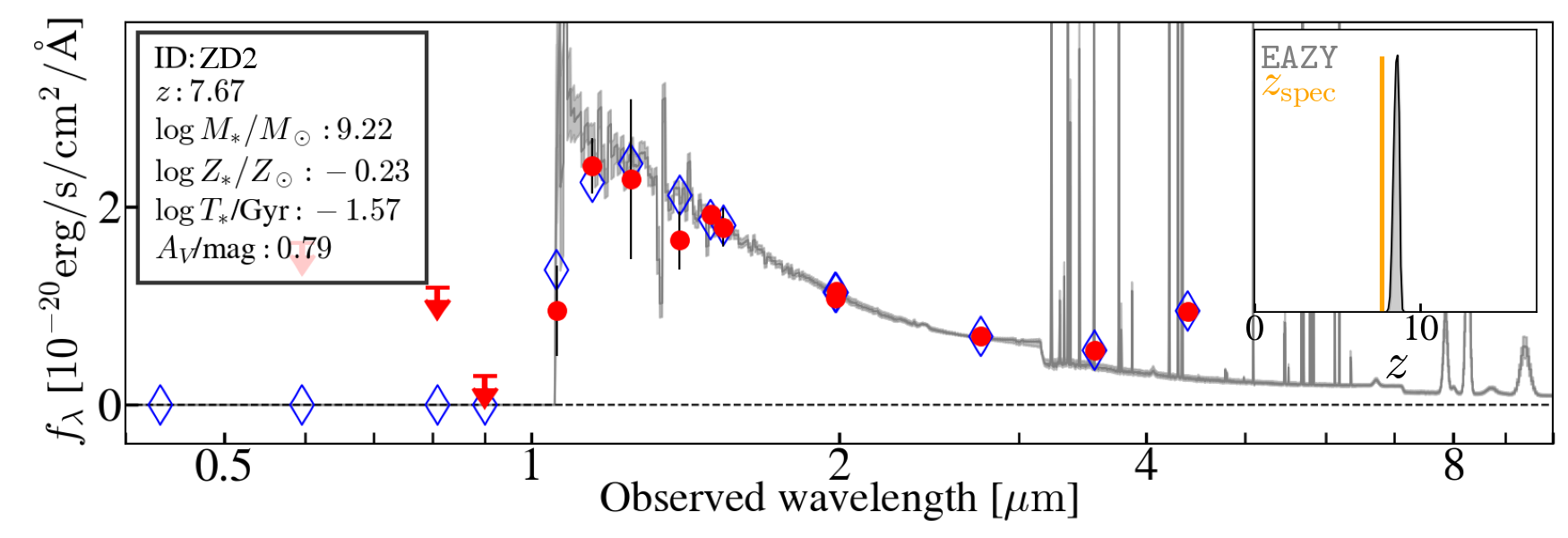}
	\includegraphics[width=.49\textwidth]{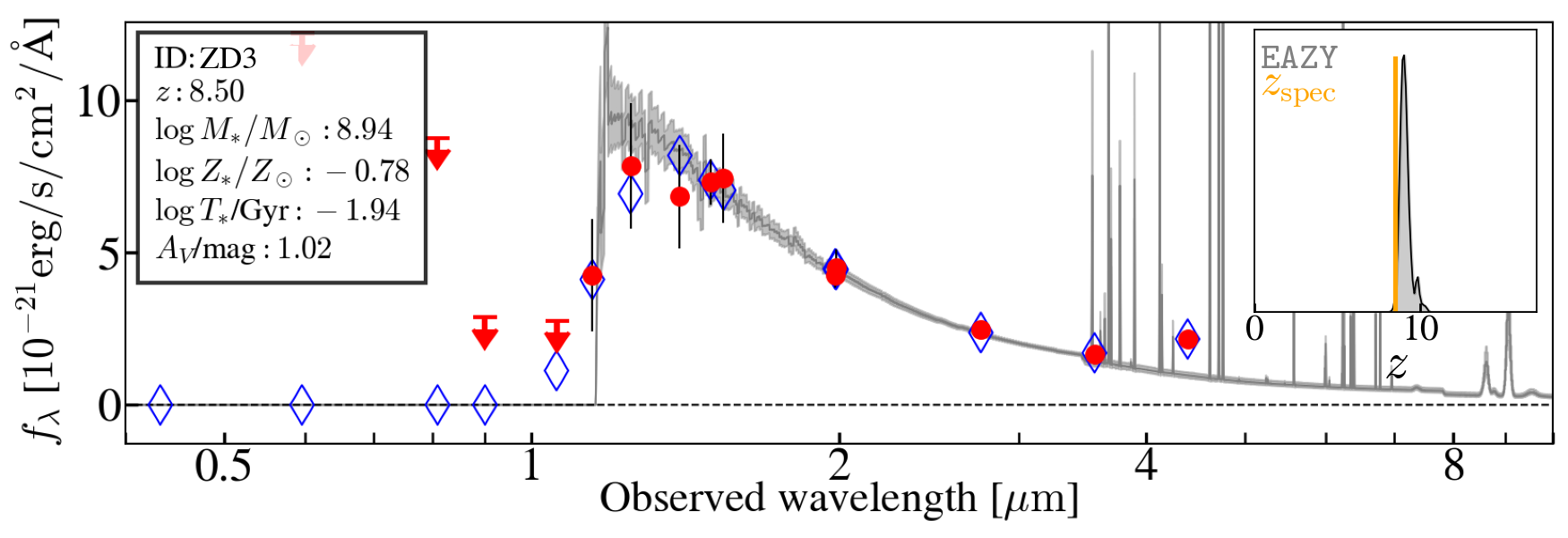}
	\includegraphics[width=.49\textwidth]{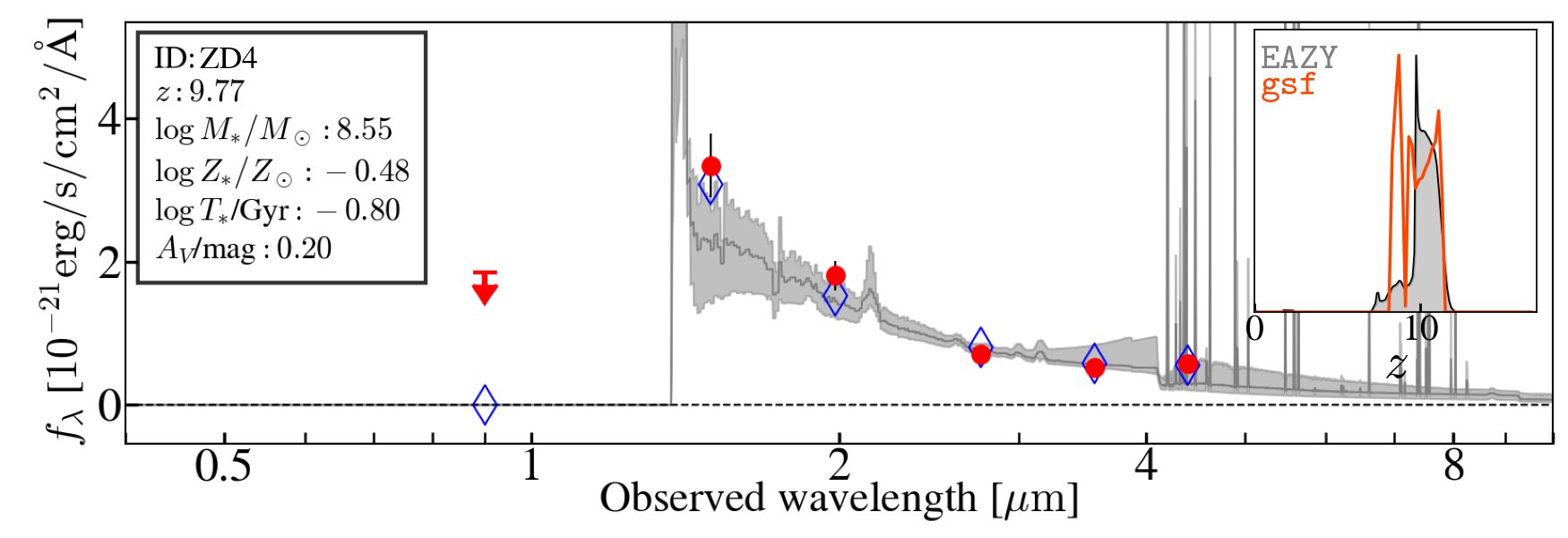}
	\includegraphics[width=0.49\textwidth]{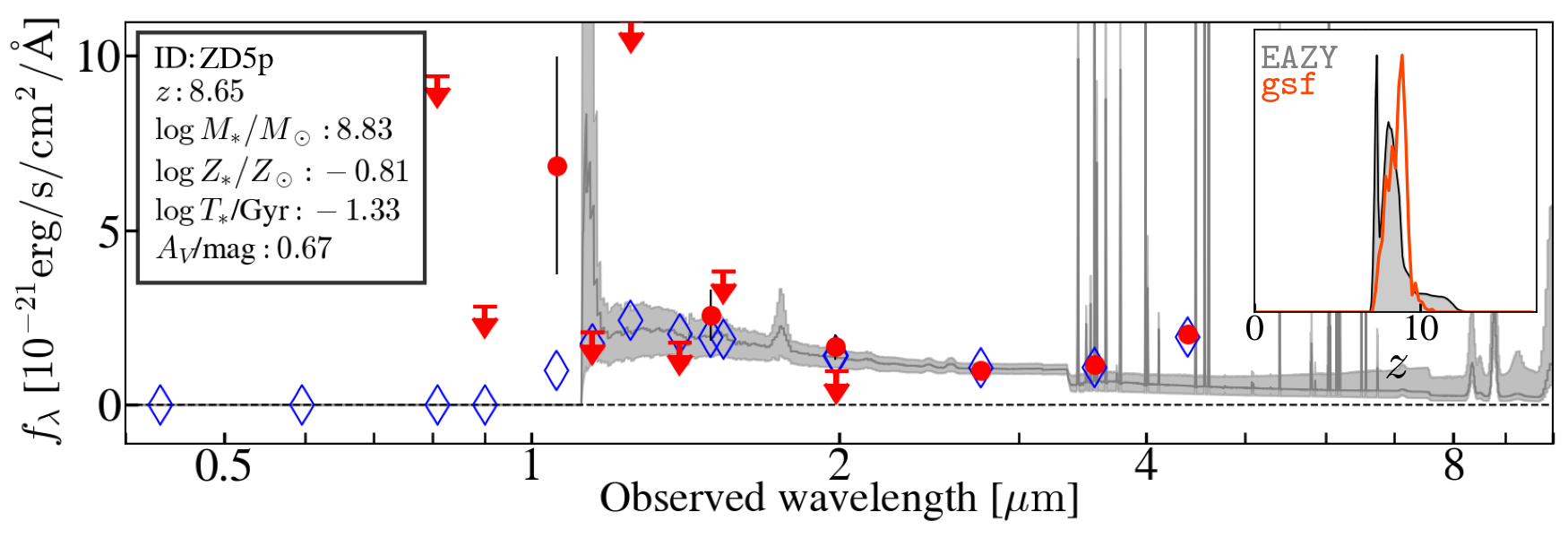}
	\includegraphics[width=0.49\textwidth]{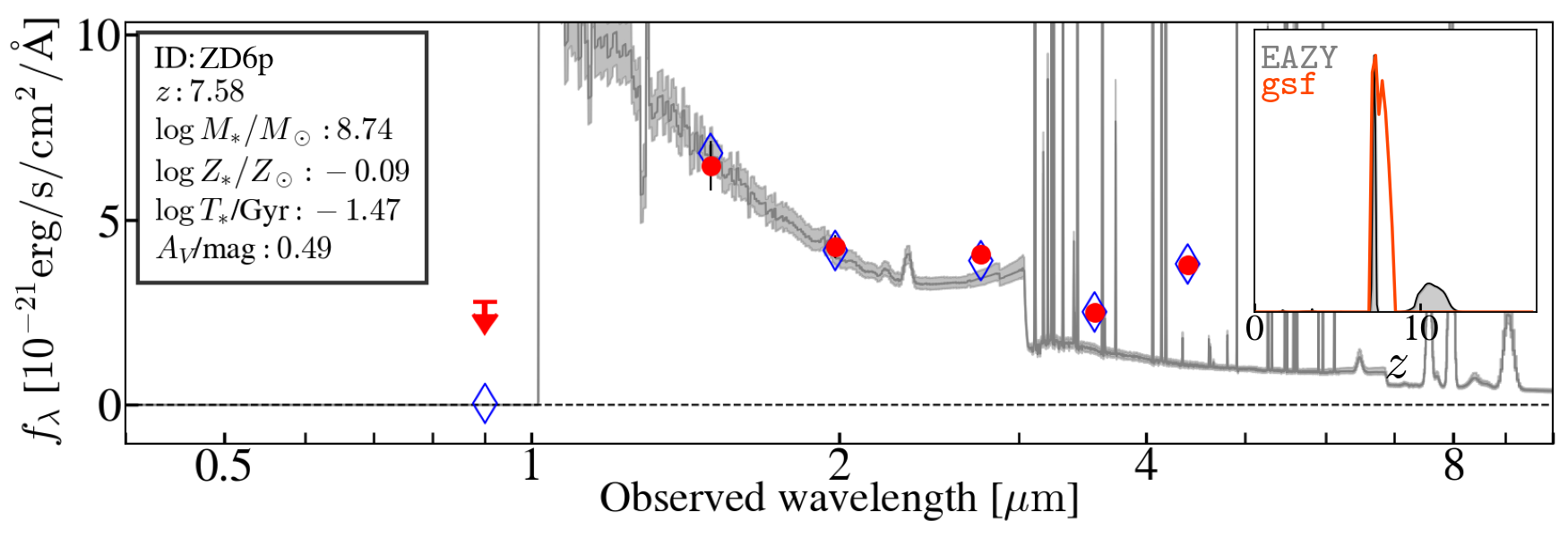}
	\includegraphics[width=0.49\textwidth]{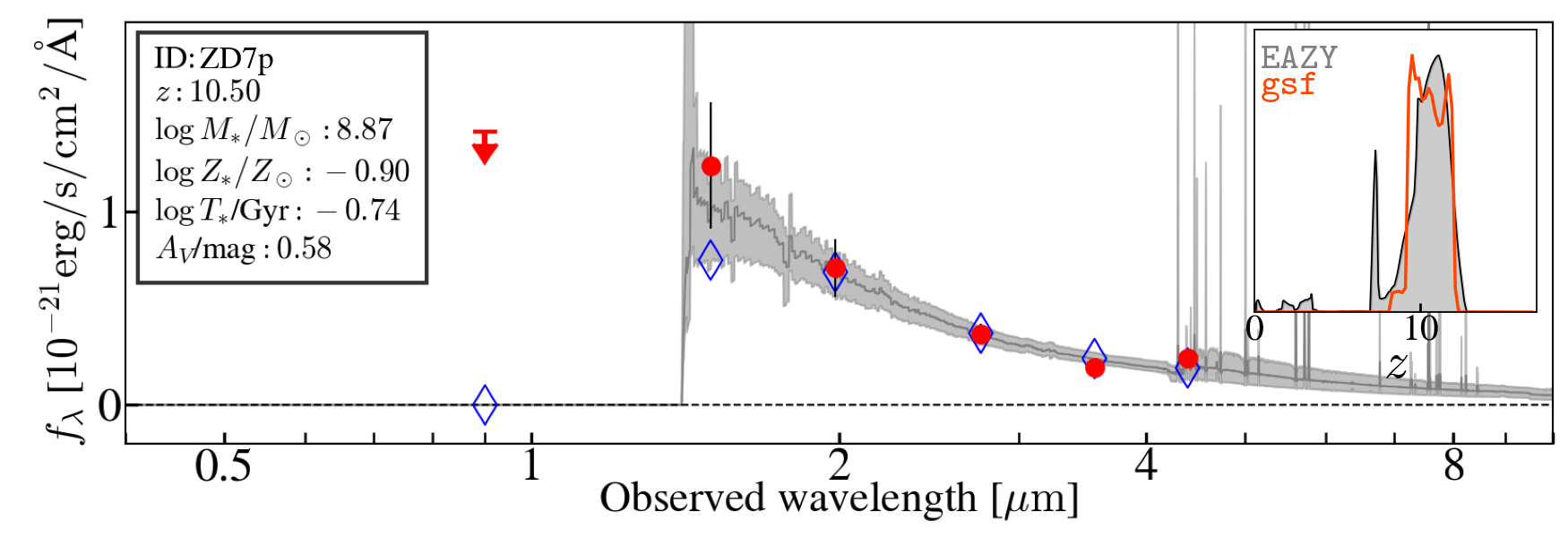}
	\includegraphics[width=0.49\textwidth]{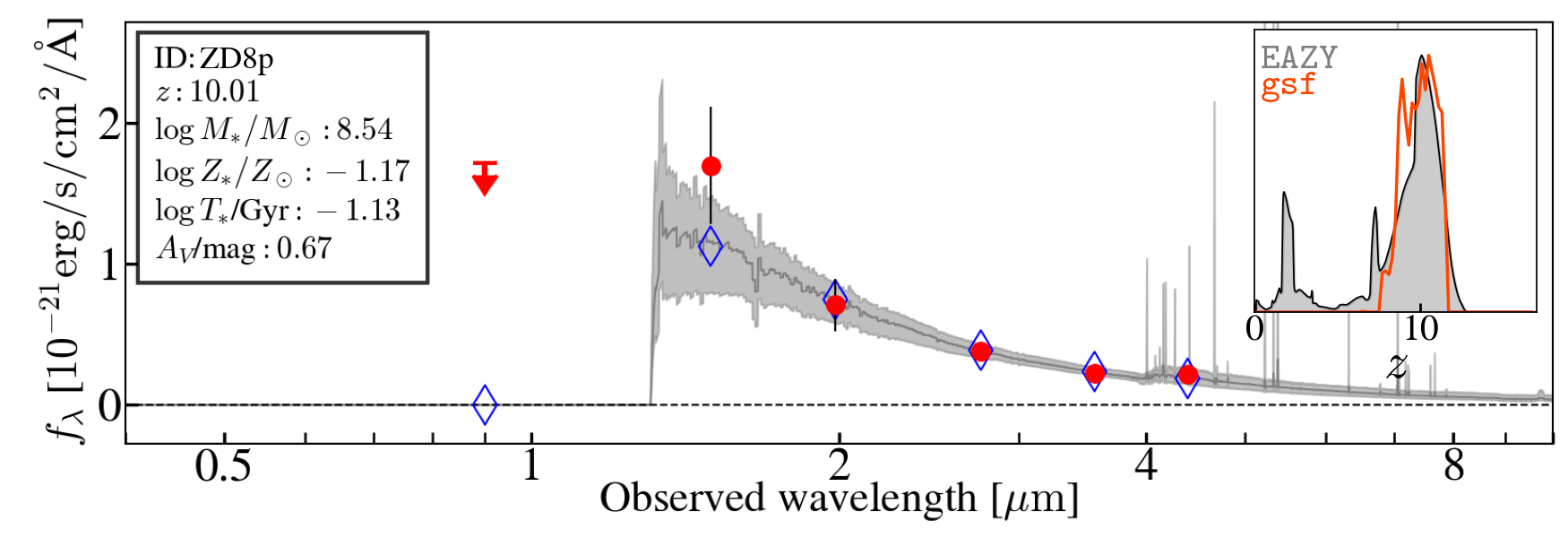}
	\includegraphics[width=.49\textwidth]{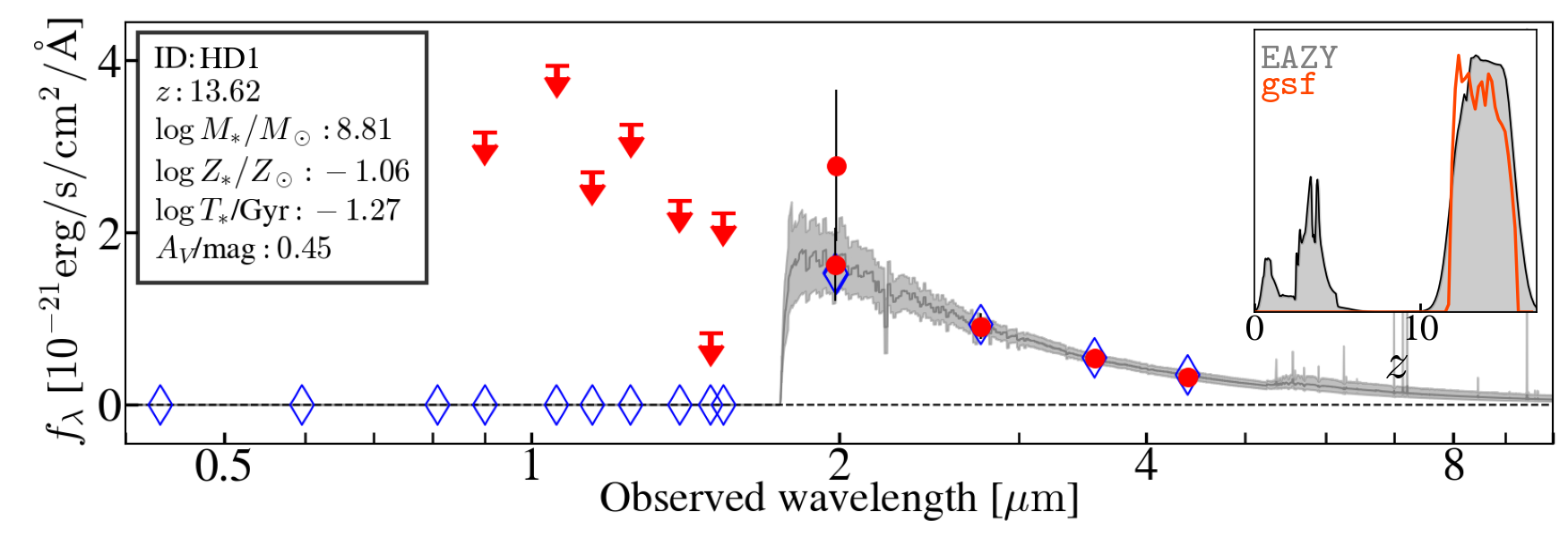}
	\includegraphics[width=.49\textwidth]{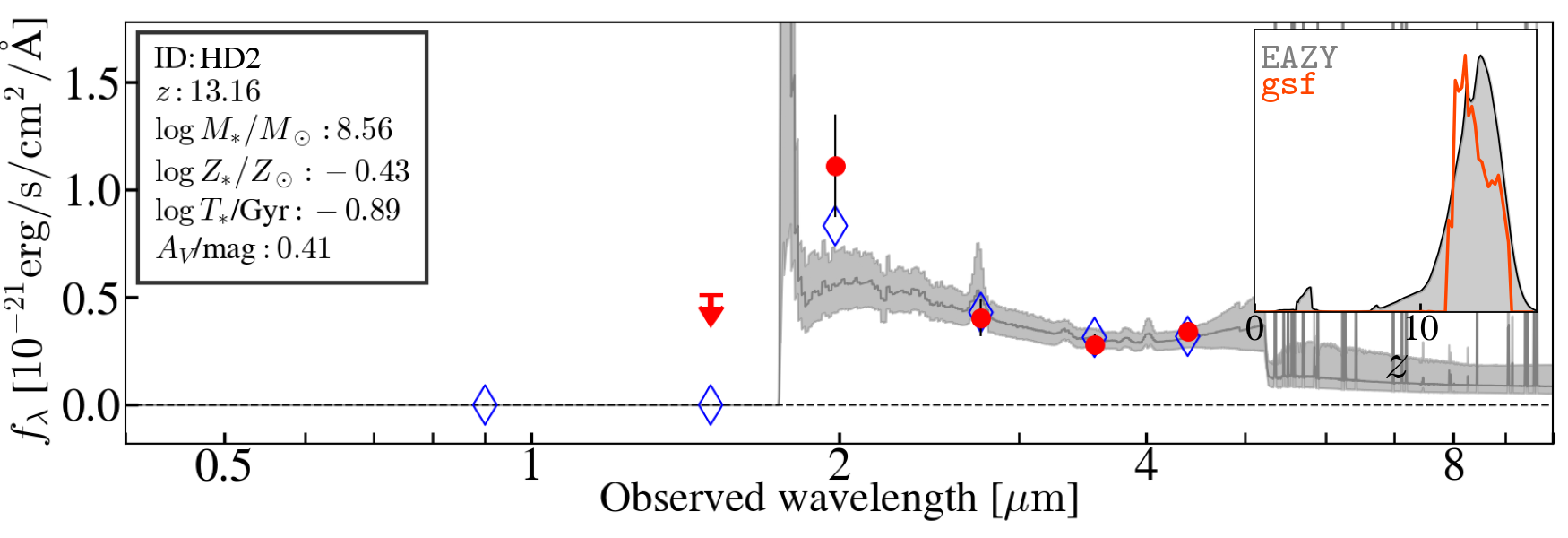}
	\includegraphics[width=0.49\textwidth]{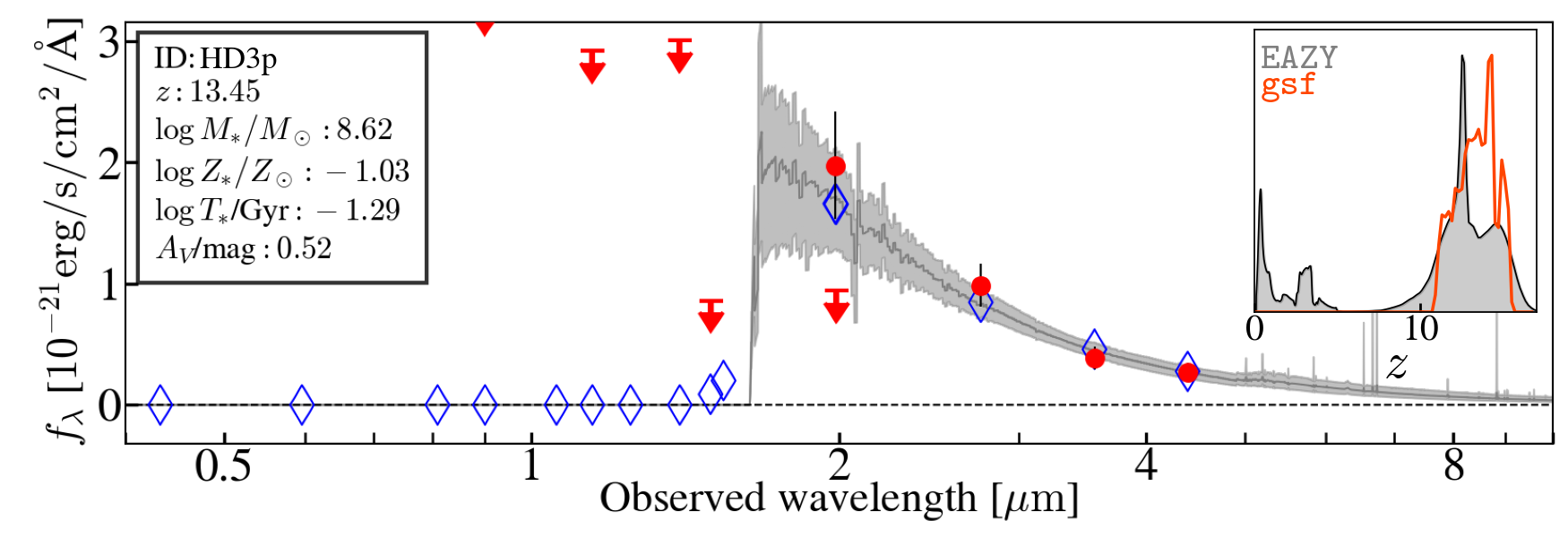}
	\caption{
	Spectral energy distribution of the final candidates. The best-fit and the 16-84th percentile range are shown (gray line and hatched region). $2\,\sigma$ error is shown along with the photometric fluxes (red symbols). Blue diamonds represent the model fluxes of the corresponding filter. Photometric redshift probability distribution (gray for \eazy\ and red for \gsf) is shown in the inset. For those with spectroscopic redshift available (yellow vertical lines in the $p(z)$ panel), redshift is fixed during the SED fitting process. It is noted that fluxes and best-fit parameters are not corrected for lens magnification in the figure.
    }
\label{fig:sed}
\end{figure*}

%=============================
\begin{deluxetable*}{ccccccccccc}%[!h]
\tabletypesize{\footnotesize}
\tablecolumns{11}
\tablewidth{0pt} 
\tablecaption{Physical properties of the final candidates.}
\tablehead{
\colhead{ID} & \colhead{$z^\dagger$} & \colhead{$M_{\rm UV}$} &\colhead{$\beta_{\lambda}$} & \colhead{$\log M_*$} & \colhead{SFR} & \colhead{$\log t_*$} & \colhead{$\log Z_*$} & \colhead{$A_V$} & \colhead{$r_e$} & \colhead{$n$}\\
\colhead{} & \colhead{} & \colhead{mag} & \colhead{} & \colhead{$M_\odot$} & \colhead{$M_\odot$\,/\,yr} & \colhead{Gyr} & \colhead{$Z_\odot$} & \colhead{mag} & \colhead{pc} & \colhead{}
}
\startdata
ZD1 & $7.663$ & $-20.05_{-0.07}^{+0.05}$ & $-2.25_{-0.05}^{+0.03}$ & $8.40_{-0.20}^{+0.18}$ & $1.15_{-0.45}^{+0.54}$ & $-1.54_{-0.69}^{+0.53}$ & $-0.43_{-0.43}^{+0.23}$ & $0.53_{-0.18}^{+0.14}$ & $119\pm7$ & $1.57\pm0.37$\\
ZD2 & $7.665$ & $-20.51_{-0.05}^{+0.05}$ & $-1.77_{-0.03}^{+0.02}$ & $9.03_{-0.14}^{+0.13}$ & $5.74_{-1.69}^{+1.72}$ & $-1.57_{-0.30}^{+0.35}$ & $-0.23_{-0.20}^{+0.07}$ & $0.79_{-0.15}^{+0.23}$ & $282\pm8$ & $0.37\pm0.06$\\
ZD3 & $8.499$ & $-17.22_{-0.09}^{+0.08}$ & $-1.88_{-0.07}^{+0.09}$ & $7.75_{-0.15}^{+0.19}$ & $0.27_{-0.07}^{+0.11}$ & $-1.94_{-0.78}^{+0.82}$ & $-0.78_{-0.37}^{+0.37}$ & $1.02_{-0.19}^{+0.13}$ & $10\pm1$ & $0.20\pm0.27$\\
ZD4 & $9.78_{-1.15}^{+1.14}$ & $-19.43_{-0.12}^{+0.12}$ & $-1.99_{-0.12}^{+0.28}$ & $8.55_{-0.90}^{+0.80}$ & $0.30_{-0.18}^{+0.66}$ & $-0.80_{-0.70}^{+0.30}$ & $-0.48_{-0.48}^{+0.59}$ & $0.20_{-0.15}^{+0.11}$ & $237\pm22$ & $0.31\pm0.27$\\
ZD5p & $8.65_{-0.72}^{+0.45}$ & $-18.19_{-0.41}^{+0.28}$ & $-1.45_{-0.26}^{+0.42}$ & $8.62_{-1.03}^{+0.51}$ & $1.14_{-1.06}^{+4.21}$ & $-1.33_{-0.58}^{+0.68}$ & $-0.81_{-0.47}^{+0.49}$ & $0.67_{-0.43}^{+1.09}$ & $184\pm12$ & $0.20\pm0.17$\\
ZD6p & $7.58_{-0.40}^{+0.43}$ & $-19.87_{-0.09}^{+0.10}$ & $-1.72_{-0.00}^{+0.06}$ & $8.74_{-0.07}^{+0.09}$ & $3.07_{-0.44}^{+0.62}$ & $-1.47_{-0.27}^{+0.23}$ & $-0.09_{-0.05}^{+0.09}$ & $0.49_{-0.08}^{+0.13}$ & $176\pm23$ & $1.55\pm0.91$\\
ZD7p & $10.50_{-0.96}^{+1.06}$ & $-18.52_{-0.26}^{+0.16}$ & $-1.89_{-0.08}^{+0.19}$ & $8.87_{-0.67}^{+0.32}$ & $0.64_{-0.47}^{+1.87}$ & $-0.74_{-0.95}^{+0.25}$ & $-0.90_{-0.74}^{+0.72}$ & $0.58_{-0.38}^{+0.60}$ & $101\pm18$ & $0.20\pm0.78$\\
ZD8p & $10.01_{-1.15}^{+0.96}$ & $-18.51_{-0.35}^{+0.22}$ & $-1.93_{-0.11}^{+0.32}$ & $8.54_{-0.42}^{+0.34}$ & $1.27_{-0.99}^{+1.39}$ & $-1.13_{-0.67}^{+0.47}$ & $-1.17_{-0.52}^{+0.96}$ & $0.67_{-0.50}^{+0.54}$ & $217\pm31$ & $0.40\pm0.44$\\
HD1 & $13.62_{-1.22}^{+1.27}$ & $-19.18_{-0.21}^{+0.19}$ & $-1.99_{-0.11}^{+0.20}$ & $8.51_{-0.40}^{+0.43}$ & $1.25_{-0.68}^{+1.70}$ & $-1.27_{-0.61}^{+0.55}$ & $-1.06_{-0.68}^{+0.91}$ & $0.45_{-0.27}^{+0.34}$ & $160\pm29$ & $0.55\pm0.37$\\
HD2 & $13.16_{-0.91}^{+1.36}$ & $-18.99_{-0.30}^{+0.22}$ & $-1.47_{-0.07}^{+0.04}$ & $8.56_{-0.86}^{+1.04}$ & $0.65_{-0.57}^{+3.48}$ & $-0.89_{-0.66}^{+0.32}$ & $-0.43_{-1.24}^{+0.59}$ & $0.41_{-0.22}^{+0.66}$ & $129\pm13$ & $0.20\pm0.50$\\
HD3p & $13.45_{-1.40}^{+1.12}$ & $-18.23_{-0.28}^{+0.19}$ & $-2.30_{-0.12}^{+0.16}$ & $7.94_{-0.45}^{+0.47}$ & $0.30_{-0.19}^{+0.49}$ & $-1.29_{-0.80}^{+0.59}$ & $-1.03_{-0.68}^{+0.96}$ & $0.52_{-0.37}^{+0.47}$ & $29\pm6$ & $0.20\pm0.76$\\
\cutinhead{Median value}
$7<z<14$ (all) & $9.78$& $-18.99$& $-1.89$& $8.55$& $1.14$& $-1.29$& $-0.78$& $0.53$& $160$& $0.31$\\
$7<z<10$ ($N=6$) & $8.08$& $-19.65$& $-1.82$& $8.59$& $1.15$& $-1.50$& $-0.45$& $0.60$& $180$& $0.34$\\
$10<z<14$ ($N=5$) & $13.16$& $-18.52$& $-1.93$& $8.54$& $0.65$& $-1.13$& $-1.03$& $0.52$& $129$& $0.20$\\
\enddata
\tablecomments{
$\dagger$: posterior redshift derived by \gsf. \ida, \idb, and \idc\ are fixed to spectroscopic redshift during the fit.
$r_e$: circularized effective radius. $n$: S\'ersic index.
$t_*$: mass-weighted age.
$Z_*$: mass-weighted metallicity.
Measurements are corrected for magnification.
}\label{tab:prop}
\end{deluxetable*}
%=============================

%The dual-module imaging provided by NIRCAM (i.e. F090W-F277W pair) rejects that this is a transient object.
% F090W : 2022-06-07T00:34:53.548
% F150W : 2022-06-07T02:56:26.340
% F200W : 2022-06-07T05:18:31.324
% F277W : 2022-06-07T00:34:53.548
% F356W : 2022-06-07T02:56:26.340
% F444W : 2022-06-07T05:18:31.324
%The presence of such closeby galaxies may affect the color of a higher-$z$  galaxy in the sightline, making an apparent dropout through attenuation by their inter-galactic medium. 

%%%%%%%%%%%%%%%%%%%%%%
\subsubsection{The phot-z sample}\label{sec:pzsample}
{ In Fig.~\ref{fig:mosaic-f090w-par}, we show six galaxies (two of which are treated as a single system; see below) identified in the phot-$z$ selection. Most of them are located near the boundary of the color selection window. Phot-$z$ samples are in general fainter than the color-cut sample; all our phot-$z$ samples have Lyman-break color (or upper limit) that is not red enough to be in the selection window.

\idg\ and \idh\ are identified in the cluster field. \idg\ is located next to a bright galaxy at $z_{\rm phot}=2.5$, showing an elongated morphology. The lens model indicates strong stretch at $\sim45$\,deg direction, which aligns with the observed elongation. It is noted that while the F150W image shows some positive pixels at the location of the object, detection is marginal ($S/N_{150}\sim1.8$). \idh\ is an isolated system but shows multiple components, clearly resolved in the NIRCam F200W stamp image. All components clearly drop out in F090W. 

In the parallel field, three galaxies are identified. Among those, IDs~949 and 963 are of particular interest sources, being located next to each other. While \sext\ deblends those into two different objects, for their proximity and similarity in color (including clear dropout in F090W), we treat the two objects (949+963; \idi) as a single system in the SED fitting analysis below. It is noted that to its northern part, there is another object (ID 793) of a similar color. We did not include this segment in our photometric analysis, because of some residual flux that appears in the dropout band. We attribute this segment to the galaxy at its west (ID~942), whose redshift is $z_{\rm phot.}\sim5$.

The remaining two phot-$z$ candidates, \idj\ and \idk, are isolated and compact. \idk\ shows a small sub-component to its west in three filters (F200W, F277W, and F356W), but this small component is not visible in F150W. None of the phot-$z$ candidates in the cluster field (ZD5p and HD3p) are listed in the RELICS catalog.
}

%%%%%%%%%%%%%%%%%%%%%
\subsection{Spectral energy distribution fitting}\label{sec:phys}
To investigate the physical properties of the selected candidates, we conduct spectral energy distribution fitting. We use SED fitting code \gsf\footnote{\url{https://github.com/mtakahiro/gsf}} \citep{morishita18,morishita19}. For those with spec-$z$, we set redshift as a fixed parameter during fitting. For the others, we set redshift as a free parameter within the 16-84\,th percentile range derived by \eazy\ (Table~\ref{tab:allcand_phot-z}). We use a template library generated by {\tt fsps} \citep{conroy09fsps}, { with the dust attenuation law derived for the SMC \citep{gordon03}}. We adopt a non-functional form for the star formation history as presented in \citet{morishita19,morishita21}, with the age pixels set to [1, 3, 10, 30, 100, 300]\,Myrs. This configuration allows flexibility in determining the star formation history and provides a clear insight into the stellar populations that consists of the observed SED. During our initial test, we found that the inclusion of emission lines is crucial to successfully fit the observed fluxes. We add an emission component to our fitting model, with two free parameters; ionization parameter set to $\log U \in [-3:-1]$, and the amplitude of this component. 

We show the results of SED analyses in Fig.~\ref{fig:sed} and summarize the derived physical parameters in Table~\ref{tab:prop}. All galaxies are in general fitted well with young stellar populations. The unprecedented coverage up to $5\,\mu$m by NIRCam reveals flux excess in the F444W filter, which is attributed to significant emission lines. For galaxies at $7<z<9$, strong emission lines such as \oiii\ doublet and \hb\ fall in the wavelength range of F444W, whereas galaxies at higher redshift show flux excess in the same filter, which is likely by strong \oii. The observed excess in the $m_{\rm F356W}-m_{\rm F444W}$ color of those galaxies clearly indicates intense star formation and hot ISM \citep[e.g.,][]{roberts-borsani20,morishita20,nicha22} and provides clues to metallicity \citep[e.g.,][]{schaerer22,tacchella22}. While direct inference of their emission properties is limited to the three spec-$z$ sources, the observed blue UV slope, $\beta_{\rm UV}$, of our sample is large in negative, { with the median value of $\sim-1.9$, which is consistent with strong line emitters found in the literature \citep{bouwens14,yamanaka19,narayakkara22}. It is also known that $\beta_{\rm UV}$ has a negative trend both with redshift and $M_{\rm UV}$ \citep{bouwens14}; however, we do not observe such a trend among our sample, likely due to the limited number of galaxies.}

\begin{figure}
\centering
	\includegraphics[width=0.48\textwidth]{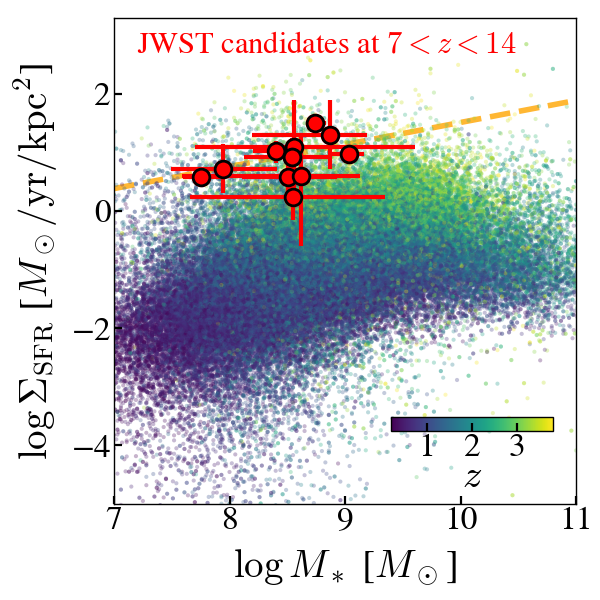}
	\caption{
    { Star formation rate surface density of our final galaxy candidates plotted as a function of stellar mass (red circles). For comparison, galaxies at $0.2<z<4$ taken from the 3DHST catalog are shown. Our sample shows a moderate correlation with stellar mass. The measured linear slope ($\sim0.38$; dashed line) is shown.}
	}
\label{fig:sfrsd}
\end{figure}

{
To further investigate the physical properties of the candidate galaxies, we derive star-formation rate surface density, $\Sigma_{\rm SFR} = {\rm SFR} / 2 \pi r_e^2$, which is a good proxy for star formation activity within the system \citep[e.g.,][]{oesch10,ono13}. We use half-light radius derived by {\tt galfit} and star-formation rate calculated by averaging the last 100\,Myr of the posterior star formation history. In Fig.~\ref{fig:sfrsd}, we show $\Sigma_{\rm SFR}$ as a function of stellar mass, along with low-$z$ galaxies taken from the 3DHST catalog \citep[][]{skelton14,vanderwel14,whitaker14}. As expected from the shape of SEDs and the flux excesses in F444W, our candidates are located at the high-locus of the distribution at the similar stellar mass. The observed increase in $\Sigma_{\rm SFR}$ compared to lower-$z$ sources is likely driven both by compact morphologies \citep[see also][]{yang22} and intense star formation at high redshifts. The observed increase in star-formation surface density indicates even stronger feedback in individual systems at the redshift range. Among our sample, a weak trend of $\Sigma_{\rm SFR}$ with stellar mass is seen, with the measured slope of $\sim0.38$. The Spearman's test indicates a moderate correlation, $R=0.54$ with $p$-value of 0.08. A followup investigation with a larger sample is of particular interest, as it may hint at the efficiency of star formation in systems of different stellar (and halo) masses.
}

%%%%%%%%%%%%%%%%%%%%%

%=============================
\begin{deluxetable}{lc}%[!h]
\tabletypesize{\footnotesize}
\tablecolumns{2}
\tablewidth{0pt}
\tablecaption{
Number densities of high-$z$ candidates.
}
\tablehead{
\colhead{$M_{\rm UV}$} & \colhead{Number density}\\
\colhead{mag} & \colhead{$\log {\rm Mpc^{-3}\,mag^{-1}}$}
}
\startdata
\cutinhead{$z\sim9$}
$-21.34$ & $<-4.13$\\
$-20.34$ & $-4.40_{-0.94}^{+0.57}$\\
$-19.34$ & $-3.87_{-0.94}^{+0.57}$\\
$-18.34$ & $-2.82_{-0.50}^{+0.37}$\\
$-17.34$ & $-3.08_{-1.64}^{+0.75}$\\
$-16.34$ & $<-2.07$\\
\cutinhead{$z\sim12$}
$-21.90$ & $<-3.92$\\
$-20.90$ & $<-3.93$\\
$-19.90$ & $<-3.89$\\
$-18.90$ & $-3.78_{-0.94}^{+0.57}$\\
$-17.90$ & $-3.70_{-1.64}^{+0.75}$\\
$-16.90$ & $<-2.63$\\
$-15.90$ & $<-2.14$\\
\cutinhead{$z\sim18$}
$-20.37$ & $<-3.85$\\
$-18.37$ & $<-3.07$\\
$-16.37$ & $<-2.16$\\
\cutinhead{$z\sim24$}
$-20.80$ & $<-3.58$\\
$-18.80$ & $<-3.11$\\
$-16.80$ & $<-1.84$\\
\enddata
\tablecomments{
$2\,\sigma$ uncertainties are quoted.
}
\label{tab:nd}
\end{deluxetable}
%=============================

%%%%%%%%%%%%%%%%%%%%%%
% New section after Ref's report;
\section{Discussion}\label{sec:disc}
{ 
In the previous sections, we present our high-$z$ galaxy candidates identified in the first deep images of JWST. Despite the limited search volume with a single pointing, we have identified 11 photometric candidates (including three spectroscopically confirmed), two of which are $z\sim13$, beyond the previous redshift limit enabled by Hubble \citep[e.g.,][]{oesch18,bouwens21}. Several teams report independent identification of high-$z$ galaxy candidates in the same field \citep{adams22,atek22,donnan22,harikane22b,yan22}. High-$z$ source selection is extremely sensitive to detailed processes during selection and the final list may vary from one to another. Therefore, it is of particular interest to investigate if our identification of 11 sources is consistent with predictions by theoretical/empirical models and if our candidates are consistent with those identified in other studies.
}

%%%%%%%%%%%%%%%%%%%%%
\begin{figure*}
\centering
	\includegraphics[width=0.4\textwidth]{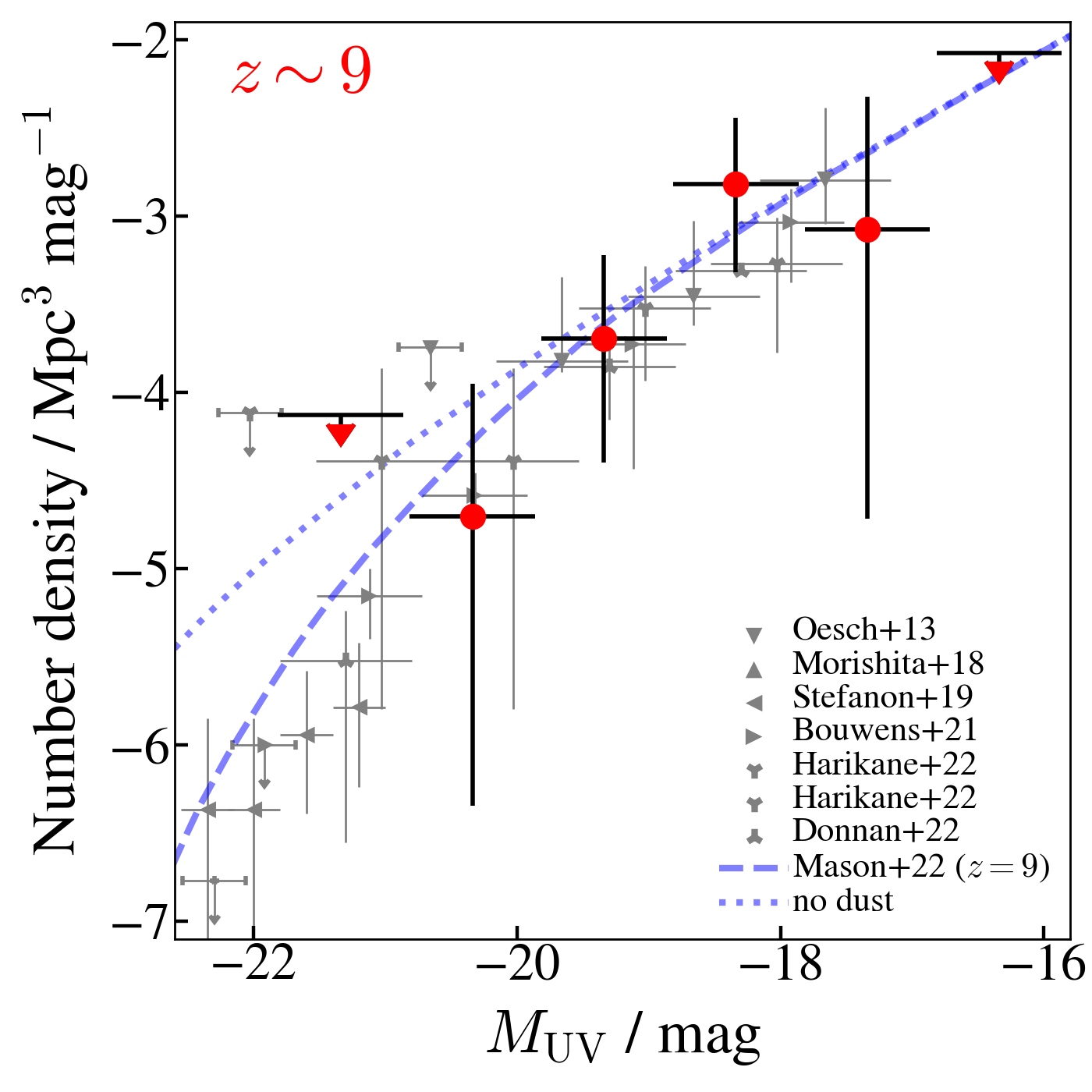}
	\includegraphics[width=0.4\textwidth]{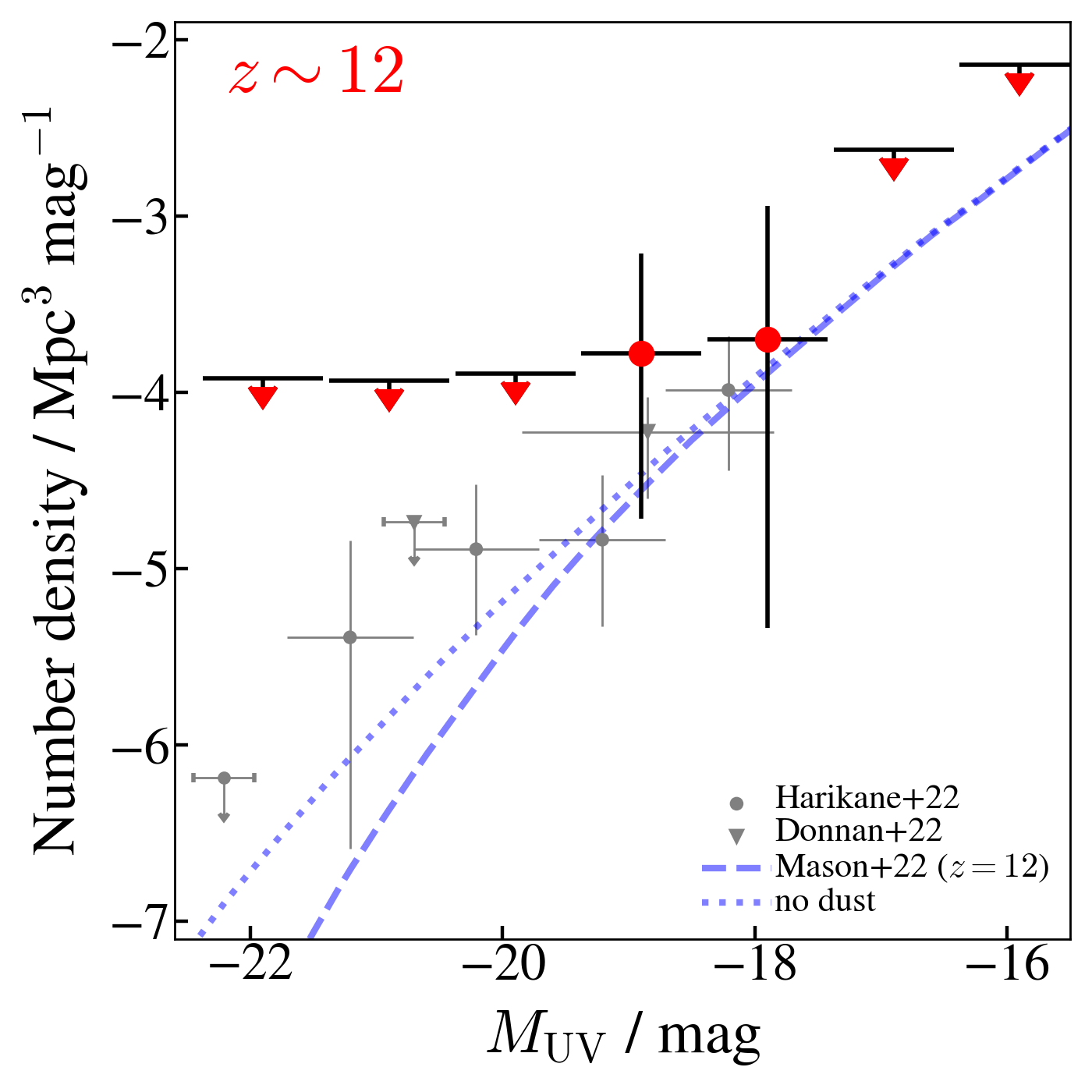}
	\includegraphics[width=0.4\textwidth]{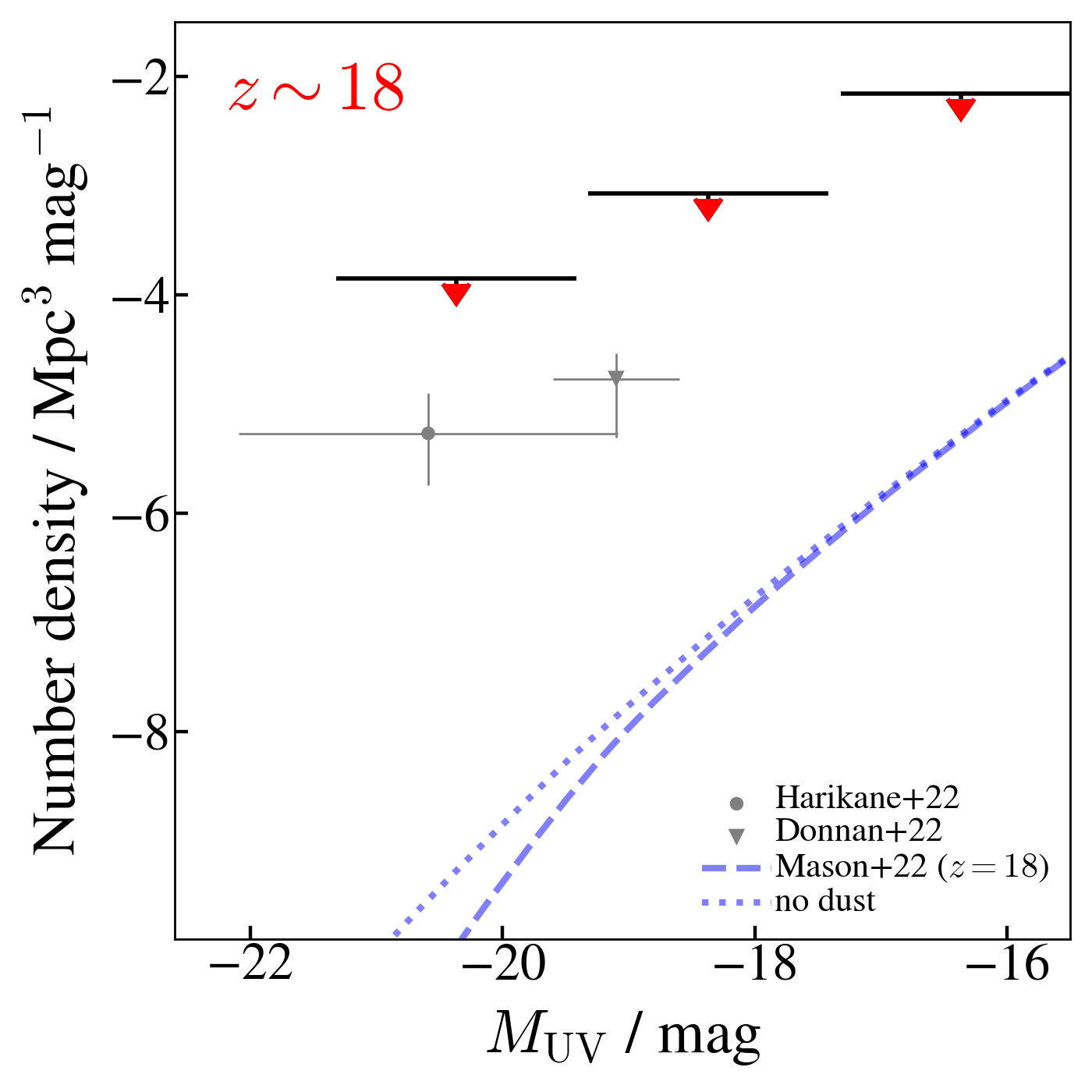}
	\includegraphics[width=0.4\textwidth]{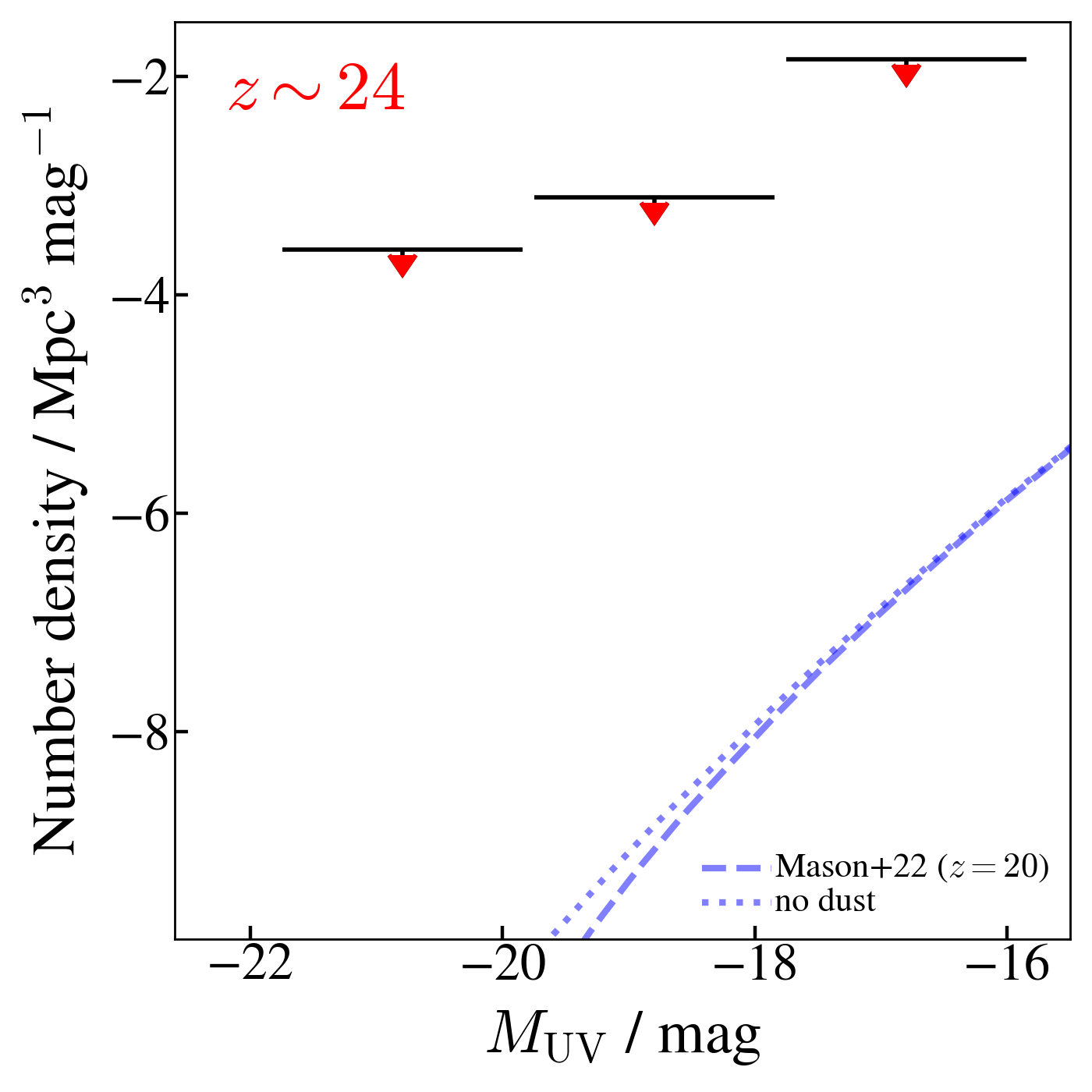}
	\caption{{ Number density estimates of the F090W-dropout ($z\sim9$, top left panel) and F150W-dropout ($z\sim12$, top right) candidates. Upper limits for the number density of F200W-dropout and F277W-dropout sources, none of which are identified here, are also shown ($z\sim18$, bottom left and $z\sim24$, bottom right). Error bars and upper limits represent $2\,\sigma$ uncertainties. Predicted luminosity functions at the corresponding redshift \citep[][dashed and dotted lines for the models with and without dust, respectively]{mason15,mason22} shown in each panel.}
	}
\label{fig:LF}
\end{figure*}

%%%%%%%%%%%%%%%%%%%%%%
{
\subsection{Number densities of high-redshift candidates}\label{sec:nd}
In Fig.~\ref{fig:LF}, we show our estimates of source number density at three redshift ranges, $z\sim9$ (F090W-dropout), $z\sim12$ (F150W-dropout), and $z\sim17$ (F200W-dropout). The number density is calculated as a function of absolute UV magnitude, $M_{\rm UV}$. The effective volume is calculated by running simulations, in the same way as in \citep{morishita18b}, for each of the color-cut selections. We calculate the detection completeness as a function of apparent F444W magnitude, $C(m)$, and the source selection function as a function of magnitude and redshift, $S(z, m)$, by analyzing artificial sources buried in the NIRCam images (and HST$+$NIRISS images in the cluster pointing). The effective comoving volume is then calculated as
\begin{equation}
V_{\rm eff}(m) = \int_{0}^{\infty} S(z,m) C(m) {dV\over{dz}} dz,
\end{equation}
where $dV/dz$ represents differential comoving volume at redshift $z$. We also corrected the volume estimate in the cluster field for magnification by using the latest version of public magnification maps, provided by Zitrin's team through the RELICS collaboration. To estimate a number density at each magnitude bin, we use the best-fit redshift and absolute UV magnitude derived by \gsf\ reported in Table~\ref{tab:prop}. The number densities of galaxies in the four redshift bins are reported in Table~\ref{tab:nd}.

Our number density estimate at $z\sim9$ is in good agreement with previous studies in the literature \citep{stefanon17,morishita18b}, as well as recent those in JWST fields \citep{donnan22,harikane22b}. We also see agreement with the empirical prediction by \citet{mason15,mason22} for the \citealt{st01} halo mass function. At $z\sim9$, the model that includes dust seems to better fit the number densities estimates here and in the literature. At higher redshift, those two models are inseparable mostly due to the lack of data points at the bright end ($M_{\rm UV}<-20$).

At $z\sim12$, where we identified three candidates, our number density estimate is slightly larger than the two JWST studies \citep{donnan22,harikane22b} and the prediction by \citet{mason22} but is still consistent within the uncertainty. It is noted that the two JWST studies explore not only the SMACS field but also CEERS (four pointings), GLASS (one pointing), and the Stephan's Quinted field \citep[only in][]{harikane22b}, resulting in $\simgt5\times$ larger volume than explored here. For the limited volume, our finding of two candidates in a single pointing may be attributed to cosmic variance. In fact, \citet{harikane22b} identify three $z\sim12$ candidates in one field while none in another field. The same is true in \citet{donnan22}, where they identify two candidates at $z\sim12$ in the SMACS field, while none in the three of the CEERS's pointings. It is also worth mentioning that that none of our $z\sim13$ candidates are identified in either of the two studies (but one, \ide, is identified by \citealt{yan22}; see below).

At $z\sim18$ and $\sim24$, where we identified no candidate, only upper limits are shown. For the volume investigated here, it is reasonable not to identify any candidates, as the prediction goes by an order of $\simgt3$ below the volume here, requiring $\sim100\times$ more volume to find one. 
}

%%%%%%%%%%%%%%%%%%%%%%
\subsection{Comparison to candidates from other studies}\label{sec:comp}
{ 
Here we investigate the consistency of the final list in multiple studies and what makes a difference by visiting all published candidates. All candidates from the literature studies are summarized in Table~\ref{tab:liter}, along with the IDs cross-matched with our catalog. It is noted that as of the writing of this manuscript, all but one of the five studies bases their analyses on the reference files in the context of {\tt jwst\_916.pmap}. As we have found in this study, the latest set of the reference files changed flux estimates for as high as $\sim40$\,\% from the original release of the data. We thus caution the readers that their future updates with the latest reference files may change their final list of high-$z$ candidates and may cause mismatch in our discussion here.

%%%
\subsubsection{Adams et al.}\label{sec:adams}
\citet{adams22} identify seven objects at $9<z<12$ via photometric redshift selection using six NIRCam filters in the SMACS0723 field. Their selection also imposes a conservative magnitude limit ($m>28.4$), which includes only candidates with strong Lyman break color. All of their objects are detected in our catalog, 5 of which remain as the final candidates in our study. Despite the difference in photometric redshift code used, their redshifts are broadly consistent with ours. The only exception is ID~1514 (\idi), whose redshift is estimated to $9.85$. While our phot-$z$ probability distribution also shows a secondary peak at $z\sim10$, we find that the best-fit template at $z=7.7$ consistently explains the observed flux `bumps' at $>2\,\mu$m, likely caused by a negative Balmer break between F277W and F356W and strong line emissions captured in F444W.

The other two objects are not selected in our final list of high-$z$ candidates, primarily because of a strong low-$z$ peak (their IDs~6115 and 10234) and suspicious positive noise in F090W (ID~10234). The SED of ID~6115 is fit well with an old galaxy template at $z\sim2.1$. The SED of ID~10234 is fit well with a strong line emitter at $z\sim3.1$, where we find that its flux excess in F200W is attributed to strong \hb+\oiii\ emission. They limit their study to $z<12$, and thus none of our three galaxies at $z>12$ appear in their study.

%%%
\subsubsection{Atek et al.}\label{sec:atek}
\citet{atek22} present two $z\sim16$ candidates, two at $z\sim12$, and seven at $9<z<11$, from color-cut selection, and four additional photo-$z$ candidates at $z\sim11$, from the SMACS0723 field. In addition to the NIRCam filters, they combined the two NIRISS filters (F115W and F200W) to improve the redshift accuracy. Among 15 of their candidates, only three are found in our final list. The majority of their objects (10) are rejected here because of significant low-$z$ probability. Despite clean non-detection in F090W, SMACS\_z16a barely misses our phot-$z$ selection ($p(z>6.5)=0.74<0.8$). The relatively high low-$z$ probability is due to the faintness of the source, resulting in its flux estimate being relatively uncertain. Use of multiple phot-$z$ estimators may recover such sources at the boundary. The other two (SMACS\_z12b, SMACS\_z16b) have suspicious positive noise in the non-detection filters for their redshift solutions (F090W and F150W, respectively).

%%%
\subsubsection{Donnan et al.}\label{sec:donnan}
\citet{donnan22} present identification of galaxies in three public fields, SMACS0723, CEERS \citep{finkelstein22}, and GLASS-ERS \citep{treu22}. From the SMACS0723 field, they identify 16 candidates at $z>8.5$ via spectral energy distribution fitting supplemented with $2\,\sigma$ non-detection in blue filters. Thus, their candidate list does not include
ZD1 and ZD2. Among their 16 candidates, four remain in our final list, including ZD3 with $z_{\rm spec}=8.5$. Nine of their candidates are excluded here due to high phot-$z$ probabilities at $z<6.5$. While one (their ID~38681) has relatively high probability, $p(z>6.5)=75\%$, its peak redshift is $z=6.7$ with a tail extending to $z<6.5$, making itself at lower $z$ compared to the galaxies in our final list. Two (38697, 12682) were excluded by our visual inspection. The former object is in a crowded region of two merging galaxies, likely a part of the merging remnants. The latter object shows a suspicious noise structure in the non-detection filter F090W. Lastly, ID~6486 is a faint object ($m_{200}=29.7$), and does not satisfy our detection limit ($S/N_{\rm det}=3.8<4$).

%%%
\subsubsection{Harikane et al.}\label{sec:harikane}
\citet{harikane22b} perform identification of galaxies in four public fields, SMACS0723, CEERS, GLASS-ERS, and the Stephan's Quintet field. They applied two color-cut selections for $z\sim12$ (F150W-dropout) and $z\sim17$ (F200W-dropout) in the SMACS0723 fields, and identify one candidate at $z\sim11.5$ and one possible candidate at $z\sim15.6$.

SM-z12-1 is detected in our catalog and satisfies non-detection criteria in F200W and bluer bands. However, the object is faint ($m_{277}=29.4$) and does not satisfy our detection limit ($S/N_{277}=2.5<3$). The Lyman break is also not large enough to satisfy our criterion (F200W$-$F277W$=1.2<1.6$). SM-z17-1 is also detected in our catalog and satisfies non-detection criteria in F200W and bluer bands. The Lyman break is not large enough to pass the limit (F200W$-$F277W$=1.2$). We note, however, that both objects have relatively high $p(z)$ values ($0.77$ and 0.78, respectively) and barely miss our phot-$z$ selection. As for SMACS\_z16a in \ref{sec:atek}, phot-$z$ estimates by another phot-$z$ code may recover those two objects as final candidates.

It is worth noting that none of our two $z\sim13$ color-cut candidates are selected in \citet{harikane22b}. While they adopt different color-cut criteria for F150W-dropout sources, we confirm that both candidates satisfy their conditions. One possible remaining factor is their additional constraint by $\Delta \chi^2 = \chi_{\rm low-z}^2-\chi_{\rm high-z}^2>4$ from their SED analyses.

%%%
\subsubsection{Yan et al.}\label{sec:yan}
\citet{yan22} present 88 objects as galaxy candidates at $11<z<20$ in the SMACS0723 field. They select their candidates via color-cut using six NIRCam filters, starting F150W as a dropout band; thus their selection does not overlap with our F090W-dropout selection. Out of their 88 sources, eight objects are found not to be listed in our parent catalog. %Those missing objects are not ...

The majority ($N=77$) of the remaining sources are not selected in this study due to high phot-$z$ probabilities at $z<6.5$. One of the objects (ID~F150DB-082) remains in our list of the phot-$z$ sample. One of the other two objects with $p(z>6.5)>0.8$, ID~F150DB-C4, shows a significantly red SED. Our SED analysis, with redshift fixed to $z=10.4$, \citet[][]{yan22}'s value, indicates a significant dust attenuation, $A_V\sim4$\,mag, which is unlikely at this redshift. We thus conclude that, despite being $p(z>6.5)=0.83$, this object is a low-$z$ interloper, likely at $z\sim4.5$, where we observe a narrow redshift peak. The other object, F200DA-034 at $z\sim19.8$, is very faint ($m_{356}=29.3$) and does not pass the detection cut ($S/N_{\rm det}=2.5<4$).
}

%%%%%%%%%%%%%%%%%%%%%%
\section{Summary}\label{sec:summary}
Already on the basis of this first data set it is clear that detections with WFC3-IR with few bands shorter than 1.6\,{\micron} are not a robust way to identify high-$z$ galaxies as at least the two sources in the cluster field had been assigned much lower redshifts. These data also allow us to begin testing the reliability of phot-$z$ at previously untested redshifts. More will be needed but it is comforting that three candidates identified in the cluster field on the basis of photometry alone were spectroscopically confirmed. This gives us hope that photometric selection is not seriously plagued by false detection. We cannot say at this time anything about its completeness given that the NIRSpec MSA slits were placed on photometrically selected objects.

We repeated the selection analyses for the cluster field with the NIRCam imaging data only. This provided the same set of color dropout galaxies selected in the first step of Sec~\ref{sec:cs}. This implies that our selection of candidates in the parallel field, despite the absence of HST coverage, is more or less of similar quality. This is also encouraging for future searches of high-$z$ sources with JWST alone (e.g., pure-parallel imaging programs).

The comparison with other studies over the same fields provides a cautionary tale on how to interpret results derived only through photometry: different studies with different but equally reasonable criteria can yield significant differences in the selected samples.

A final note is that our analysis may still not be complete in terms of source detection. The (in)famous hexagonal PSF spikes from bright stars severely contaminate surrounding pixels and make it challenging to detect all background sources. In addition, the cluster field is even more challenging, being affected by brightest cluster galaxies. A dedicated analysis, such as one presented in \citet[][]{bouwens21}, will be required for detection completeness in crowded fields. 

\targ\ is not only the deepest NIR imaging field at this time, but also has rich spectroscopic data obtained by NIRSPEC and NIRISS as well as mid-IR imaging by MIRI collected during the ERO campaign. Future follow-up studies with those data will further reveal the nature of the three galaxies in the cluster field. While the additional candidates newly identified in this study are not as exceptionally bright as those in the cluster field, they are still within the reach of sensitive NIRSPEC spectroscopy for its strong emission lines revealed in our study. For future spectroscopic study and planning, we publish the best-fit templates of our final galaxies on a dedicated website.\footnote{\url{https://github.com/mtakahiro/SMACS0723_highz_SED_templates}}

%=============================
\section*{Acknowledgements}
%Support for program JWST-{r xxx} was provided by NASA through a grant from the Space Telescope   Science Institute, which is operated by the Association of Universities for Research in Astronomy, Inc., under NASA contract NAS 5-03127. 
We would like to thank the anonymous referee for providing extremely valuable comments, which improved the manuscript significantly. The Early Release Observations and associated materials were developed, executed, and compiled by the ERO production team:  Hannah Braun, Claire Blome, Matthew Brown, Margaret Carruthers, Dan Coe, Joseph DePasquale, N\'estor Espinoza, Macarena Garcia Marin, Karl Gordon, Alaina Henry, Leah Hustak, Andi James, Ann Jenkins, Anton Koekemoer, Stephanie LaMassa, David Law, Alexandra Lockwood, Amaya Moro-Martin, Susan Mullally, Alyssa Pagan, Dani Player, Klaus Pontoppidan, Charles Proffitt, Christine Pulliam, Leah Ramsay, Swara Ravindranath, Neill Reid, Massimo Robberto, Elena Sabbi, Leonardo Ubeda. The EROs were also made possible by the foundational efforts and support from the JWST instruments, STScI planning and scheduling, and Data Management teams.
Some of the data presented in this paper were obtained from the Mikulski Archive for Space Telescopes (MAST) at the Space Telescope Science Institute. The specific observations analyzed can be accessed via \dataset[10.17909/67ft-nb86]{https://doi.org/10.17909/67ft-nb86}. TM is grateful to Guido Roberts-Borsani, Benedetta Vulcani, and Xin Wang for our constructive discussion at UCLA and their insights into the new dataset from this brand-new observatory, and to Tommaso Treu for his kind and generous support. 

\software{\eazy\ \citep{brammer08}, \sext\ \citep[v2.25.0,][]{bertin96}, gsf \citep{morishita18,morishita19}, fsps \citep{conroy09fsps}}

%%%%%%%%%%%%%
% Comparison to literature;
%%%%%%%%%%%%%
\begin{deluxetable*}{cccclccl}
\tablecaption{{ List of high-$z$ candidates identified in the literature}}
\tabletypesize{\footnotesize}
\tablecolumns{8}
\tablewidth{0pt} 
\tablehead{
\colhead{ID$_{\rm liter.}$} & \colhead{R.A.} & \colhead{Decl.} & \colhead{$z_{\rm liter.}$} & \colhead{ID} & \colhead{$z$} & \colhead{$p(z>6.5)$} & \colhead{Comments}\\
\colhead{} & \colhead{degree} & \colhead{degree} & \colhead{} & \colhead{} & \colhead{} & \colhead{} & \colhead{}
}
\startdata
\cutinhead{Adams et al.}
1696 & 1.1083435e+02 & -7.3434509e+01 & 9.59 & WDF-C-769 & $7.34_{-0.07}^{+0.05}$ & 1.00 & \\
2462 & 1.1084492e+02 & -7.3435043e+01 & 9.50 & WDF-C-1045 & $8.55_{-0.18}^{+0.15}$ & 1.00 & \\
6878 & 1.1085982e+02 & -7.3449127e+01 & 9.59 & WDF-C-3186 & $9.03_{-0.25}^{+0.31}$ & 1.00 & \\
1514 & 1.1061509e+02 & -7.3477554e+01 & 9.85 & WDF-P-963 & $7.35_{-0.18}^{+3.65}$ & 0.99 &\\
2779 & 1.1064618e+02 & -7.3475845e+01 & 9.51 & WDF-P-1762 & $10.26_{-0.66}^{+0.72}$ & 1.00 &\\
6115 & 1.1071698e+02 & -7.3464973e+01 & 10.94 & WDF-P-2289 & $2.05_{-1.12}^{+8.04}$ & 0.17 & $p(z)$\\
10234 & 1.1066535e+02 & -7.3501740e+01 & 11.42 & WDF-P-6576 & $3.05_{-2.81}^{+0.27}$ & 0.03 & $p(z)$. Suspicious noise in F090W.\\
\cutinhead{Atek et al.}
SMACS\_z10a & 1.1085982e+02 & -7.3449127e+01 & 9.92 & WDF-C-3186 & $9.03_{-0.25}^{+0.31}$ & 1.00 & \\
SMACS\_z10b & 1.1084492e+02 & -7.3435043e+01 & 9.79 & WDF-C-1045 & $8.55_{-0.18}^{+0.15}$ & 1.00 & \\
SMACS\_z10c & 1.1083435e+02 & -7.3434509e+01 & 9.94 & WDF-C-769 & $7.34_{-0.07}^{+0.05}$ & 1.00 & \\
SMACS\_z11c & 1.1075713e+02 & -7.3448654e+01 & 11.22 & WDF-C-629 & $3.02_{-1.56}^{+7.79}$ & 0.34 & $p(z)$\\
SMACS\_z16a & 1.1086060e+02 & -7.3467911e+01 & 15.97 & WDF-C-5261 & $9.84_{-7.33}^{+1.38}$ & 0.74 & $p(z)$.\\ %{r check Dwarf}\\
SMACS\_z10d & 1.1069465e+02 & -7.3478012e+01 & 9.98 & WDF-P-3632 & $2.34_{-1.27}^{+2.30}$ & 0.07 & $p(z)$\\
SMACS\_z10e & 1.1068899e+02 & -7.3491814e+01 & 10.44 & WDF-P-6786 & $4.62_{-2.91}^{+6.02}$ & 0.32 & $p(z)$\\
SMACS\_z10f & 1.1073891e+02 & -7.3487900e+01 & 10.47 & WDF-P-6546 & $1.92_{-1.01}^{+1.07}$ & 0.02 & $p(z)$\\
SMACS\_z11a & 1.1066482e+02 & -7.3494514e+01 & 10.75 & WDF-P-5279 & $3.70_{-1.73}^{+7.78}$ & 0.34 & $p(z)$\\
SMACS\_z11b & 1.1072444e+02 & -7.3473129e+01 & 11.22 & WDF-P-3753 & $1.99_{-0.65}^{+0.93}$ & 0.01 & $p(z)$\\
SMACS\_z11d & 1.1065319e+02 & -7.3469231e+01 & 11.28 & WDF-P-893 & $0.72_{-0.40}^{+1.56}$ & 0.00 & $p(z)$\\
SMACS\_z11e & 1.1070534e+02 & -7.3462379e+01 & 11.52 & WDF-P-1486 & $3.51_{-2.07}^{+7.80}$ & 0.24 & $p(z)$\\
SMACS\_z12a & 1.1069765e+02 & -7.3500481e+01 & 12.03 & WDF-P-5896 & $12.68_{-10.04}^{+1.91}$ & 0.68 & $p(z)$\\
SMACS\_z12b & 1.1071779e+02 & -7.3465393e+01 & 12.35 & WDF-P-2356 & $13.83_{-9.35}^{+1.72}$ & 0.82 & suspicious noise in F090W\\
SMACS\_z16b & 1.1066457e+02 & -7.3502274e+01 & 15.70 & WDF-P-6055 & $15.95_{-2.15}^{+0.89}$ & 0.91 & detection in F150W\\
\cutinhead{Donnan et al.}
38681 & 1.1086747e+02 & -7.3438889e+01 & 8.57 & WDF-C-2143 & $7.31_{-2.21}^{+0.91}$ & 0.75 & $p(z)$. $z_{\rm peak}\sim6.7$\\ % F090W$-$F150W$=1.1$\\
39556 & 1.1086165e+02 & -7.3436226e+01 & 8.86 & WDF-C-1622 & $8.20_{-0.78}^{+1.03}$ & 1.00 & \\
28093 & 1.1090845e+02 & -7.3455986e+01 & 9.16 & WDF-C-4967 & $2.78_{-2.11}^{+7.26}$ & 0.32 & $p(z)$\\
34086 & 1.1085982e+02 & -7.3449127e+01 & 9.36 & WDF-C-3186 & $9.03_{-0.25}^{+0.31}$ & 1.00 & \\
38697 & 1.1086647e+02 & -7.3438789e+01 & 9.36 & WDF-C-2064 & $8.50_{-1.11}^{+1.18}$ & 0.98 & merger remnant?\\
35470 & 1.1076282e+02 & -7.3446548e+01 & 12.03 & WDF-C-512 & $9.47_{-7.31}^{+3.57}$ & 0.64 & $p(z)$\\
40079 & 1.1080821e+02 & -7.3434746e+01 & 14.28 & WDF-C-237 & $5.54_{-3.17}^{+7.50}$ & 0.45 & $p(z)$\\
9544 & 1.1066106e+02 & -7.3479889e+01 & 9.06 & WDF-P-2590 & $2.06_{-0.26}^{+0.29}$ & 0.00 & $p(z)$\\
12682 & 1.1066240e+02 & -7.3475121e+01 & 9.47 & WDF-P-2194 & $9.44_{-2.37}^{+1.56}$ & 0.86 & Suspicious noise in F090W. \\%F090W-F150W=1.0. 
22480 & 1.1069090e+02 & -7.3462929e+01 & 9.47 & WDF-P-1095 & $10.53_{-1.55}^{+1.00}$ & 0.97 & \\
12218 & 1.1064618e+02 & -7.3475845e+01 & 9.68 & WDF-P-1762 & $10.26_{-0.66}^{+0.72}$ & 1.00 & \\
15019 & 1.1074287e+02 & -7.3472061e+01 & 9.68 & WDF-P-4203 & $7.28_{-5.41}^{+3.42}$ & 0.53 & $p(z)$\\
3763 & 1.1070474e+02 & -7.3492020e+01 & 9.78 & WDF-P-6064 & $3.51_{-2.94}^{+8.19}$ & 0.24 & $p(z)$\\
6200 & 1.1067314e+02 & -7.3486275e+01 & 9.78 & WDF-P-4247 & $8.28_{-6.05}^{+3.05}$ & 0.60 & $p(z)$\\
21901 & 1.1069484e+02 & -7.3463715e+01 & 12.16 & WDF-P-1363 & $3.16_{-0.56}^{+1.06}$ & 0.15 & $p(z)$\\
6486 & 1.1072211e+02 & -7.3485748e+01 & 12.56 & WDF-P-6773 & $14.11_{-9.39}^{+1.89}$ & 0.81 & $S/N_{\rm det}=3.8<4$\\
\cutinhead{Harikane et al.}
SM-z12-1 & 1.1070145e+02 & -7.3476830e+01 & 11.52 & WDF-P-3651 & $12.20_{-7.58}^{+7.25}$ & 0.77 & $S/N_{\rm F277W}=2.5$, F200W$-$F277W$=1.2$\\
SM-z17-1$^\dagger$ & 1.1087592e+02 & -7.3458427e+01 & 15.63 & WDF-C-5499 & $14.73_{-10.57}^{+4.52}$ & 0.78 & F200W$-$F277W$=1.2$\\
\enddata
\tablecomments{
Sources that remained in our final list are listed without comments.
$\dagger$: Listed as a potential candidate.
}\label{tab:liter}
\end{deluxetable*}

\begin{deluxetable*}{cccclccl}
\tablecaption{{ (Continued) List of high-$z$ candidates identified in the literature}}
\tabletypesize{\footnotesize}
\tablecolumns{8}
\tablewidth{0pt} 
\tablehead{
\colhead{ID$_{\rm liter.}$} & \colhead{R.A.} & \colhead{Decl.} & \colhead{$z_{\rm liter.}$} & \colhead{ID} & \colhead{$z$} & \colhead{$p(z>6.5)$} & \colhead{Comments}\\
\colhead{} & \colhead{degree} & \colhead{degree} & \colhead{} & \colhead{} & \colhead{} & \colhead{} & \colhead{}
}
\startdata
\cutinhead{Yan et al.$^\dagger$}
F150DB-004 & 1.1081004e+02 & -7.3469292e+01 & 14.00 & WDF-C-5150 & $5.61_{-0.22}^{+0.24}$ & 0.00 & $p(z)$\\
F150DB-007 & 1.1085056e+02 & -7.3466286e+01 & 14.60 & WDF-C-5204 & $10.41_{-9.35}^{+2.20}$ & 0.59 & $p(z)$\\
F150DB-011 & 1.1086408e+02 & -7.3466446e+01 & 11.60 & WDF-C-5176 & $1.52_{-1.06}^{+3.22}$ & 0.06 & $p(z)$\\
F150DB-013 & 1.1077367e+02 & -7.3464104e+01 & 11.40 & WDF-C-2937 & $2.90_{-2.67}^{+0.39}$ & 0.00 & $p(z)$\\
F150DB-021 & 1.1080333e+02 & -7.3462578e+01 & 11.80 & WDF-C-3474 & $2.84_{-1.14}^{+5.61}$ & 0.44 & $p(z)$\\
F150DB-023 & 1.1077446e+02 & -7.3462074e+01 & 7.20 & WDF-C-2724 & $6.07_{-4.23}^{+3.95}$ & 0.46 & $p(z)$\\
F150DB-026 & 1.1084926e+02 & -7.3461319e+01 & 11.40 & WDF-C-4828 & $2.91_{-2.57}^{+0.45}$ & 0.00 & $p(z)$\\
F150DB-031 & 1.1083992e+02 & -7.3460083e+01 & 11.60 & WDF-C-4061 & $3.90_{-2.46}^{+7.10}$ & 0.25 & $p(z)$\\
F150DB-033 & 1.1087781e+02 & -7.3459175e+01 & 14.80 & WDF-C-5690 & $3.36_{-1.89}^{+10.80}$ & 0.23 & $p(z)$\\
F150DB-040 & 1.1080025e+02 & -7.3456947e+01 & 10.80 & WDF-C-2728 & $2.80_{-2.59}^{+0.39}$ & 0.01 & $p(z)$\\
F150DB-041 & 1.1077812e+02 & -7.3457100e+01 & 16.00 & WDF-C-2145 & $0.53_{-0.15}^{+3.42}$ & 0.03 & $p(z)$\\
F150DB-044 & 1.1091432e+02 & -7.3456139e+01 & 11.60 & WDF-C-5213 & $2.61_{-2.36}^{+0.67}$ & 0.01 & $p(z)$\\
F150DB-048 & 1.1075723e+02 & -7.3455078e+01 & 15.00 & WDF-C-1268 & $6.12_{-3.90}^{+6.47}$ & 0.46 & $p(z)$\\
F150DB-050 & 1.1085293e+02 & -7.3454163e+01 & 11.60 & WDF-C-3629 & $2.93_{-2.66}^{+0.80}$ & 0.01 & $p(z)$\\
F150DB-052 & 1.1086780e+02 & -7.3453796e+01 & 15.00 & WDF-C-3964 & $12.42_{-9.49}^{+2.01}$ & 0.73 & $p(z)$\\
F150DB-054 & 1.1080293e+02 & -7.3452637e+01 & 11.40 & WDF-C-2235 & $0.48_{-0.19}^{+0.08}$ & 0.00 & $p(z)$\\
F150DB-056 & 1.1078082e+02 & -7.3452866e+01 & 7.20 & WDF-C-1658 & $11.53_{-8.51}^{+3.21}$ & 0.72 & $p(z)$\\
F150DB-058 & 1.1085042e+02 & -7.3452530e+01 & 15.20 & WDF-C-3407 & $0.27_{-0.14}^{+0.12}$ & 0.00 & $p(z)$\\
F150DB-069 & 1.1076826e+02 & -7.3448433e+01 & 11.80 & WDF-C-865 & $12.48_{-9.39}^{+1.90}$ & 0.77 & $p(z)$\\
F150DB-075 & 1.1075976e+02 & -7.3444901e+01 & 11.40 & WDF-C-284 & $2.89_{-2.59}^{+0.31}$ & 0.00 & $p(z)$\\
F150DB-076 & 1.1087300e+02 & -7.3444359e+01 & 11.60 & WDF-C-2959 & $0.93_{-0.47}^{+2.77}$ & 0.00 & $p(z)$\\
F150DB-079 & 1.1080526e+02 & -7.3441589e+01 & 13.80 & WDF-C-931 & $3.36_{-0.79}^{+9.25}$ & 0.38 & $p(z)$\\
F150DB-082 & 1.1084528e+02 & -7.3440430e+01 & 11.60 & WDF-C-1730 & $13.39_{-9.37}^{+1.65}$ & 0.81 & \\
F150DB-084 & 1.1078188e+02 & -7.3439972e+01 & 11.60 & WDF-C-226 & $4.00_{-3.33}^{+0.92}$ & 0.10 & $p(z)$\\
F150DB-088 & 1.1080888e+02 & -7.3438148e+01 & 11.60 & WDF-C-626 & $3.58_{-1.28}^{+8.92}$ & 0.34 & $p(z)$\\
F150DB-090 & 1.1085972e+02 & -7.3437157e+01 & 11.40 & WDF-C-1643 & $2.95_{-1.16}^{+0.68}$ & 0.00 & $p(z)$\\
F150DB-095 & 1.1085355e+02 & -7.3433678e+01 & 11.60 & WDF-C-1144 & $2.77_{-2.01}^{+1.86}$ & 0.07 & $p(z)$\\
F150DB-C4 & 1.1085859e+02 & -7.3444389e+01 & 10.40 & WDF-C-2420 & $15.69_{-10.25}^{+1.10}$ & 0.83 & Likely a dusty interloper at $z\sim4.5$.\\
F200DB-015 & 1.1078262e+02 & -7.3467125e+01 & 16.00 & WDF-C-3520 & $9.68_{-6.40}^{+10.66}$ & 0.62 & $p(z)$\\
F200DB-045 & 1.1084516e+02 & -7.3461304e+01 & 20.40 & WDF-C-4322 & $0.40_{-0.26}^{+0.15}$ & 0.00 & $p(z)$\\
F200DB-086 & 1.1077728e+02 & -7.3455536e+01 & 15.40 & WDF-C-1932 & $14.60_{-9.97}^{+5.76}$ & 0.76 & $p(z)$\\
F200DB-109 & 1.1090538e+02 & -7.3453957e+01 & 15.80 & WDF-C-5512 & $0.05_{-0.03}^{+0.04}$ & 0.00 & $p(z)$\\
F200DB-159 & 1.1085593e+02 & -7.3445839e+01 & 16.00 & WDF-C-2731 & $1.97_{-1.59}^{+0.32}$ & 0.01 & $p(z)$\\
F200DB-175 & 1.1079664e+02 & -7.3443909e+01 & 16.20 & WDF-C-999 & $5.07_{-3.46}^{+4.32}$ & 0.34 & $p(z)$\\
F200DB-181 & 1.1080309e+02 & -7.3442154e+01 & 15.80 & WDF-C-946 & $4.21_{-2.69}^{+1.03}$ & 0.13 & $p(z)$\\
F277DB-001 & 1.1082343e+02 & -7.3473969e+01 & -99.00 & WDF-C-5008 & $7.66_{-5.05}^{+5.95}$ & 0.55 & $p(z)$\\
\enddata
\tablecomments{
$\dagger$: 80 objects that are detected in our catalog, out of 88 in the original study by \citet{yan22}, are shown.
}\label{tab:liter2}
\end{deluxetable*}

\begin{deluxetable*}{cccclccl}
\tablecaption{{ (Continued) List of high-$z$ candidates identified in the literature}}
\tabletypesize{\footnotesize}
\tablecolumns{8}
\tablewidth{0pt} 
\tablehead{
\colhead{ID$_{\rm liter.}$} & \colhead{R.A.} & \colhead{Decl.} & \colhead{$z_{\rm liter.}$} & \colhead{ID} & \colhead{$z$} & \colhead{$p(z>6.5)$} & \colhead{Comments}\\
\colhead{} & \colhead{degree} & \colhead{degree} & \colhead{} & \colhead{} & \colhead{} & \colhead{} & \colhead{}
}
\startdata
\cutinhead{Yan et al.}
F150DB-013 & 1.1073157e+02 & -7.3465691e+01 & 11.40 & WDF-P-2859 & $1.64_{-1.41}^{+1.64}$ & 0.01 & $p(z)$\\
F150DA-008 & 1.1067804e+02 & -7.3498581e+01 & 13.40 & WDF-P-6060 & $4.07_{-2.67}^{+3.78}$ & 0.20 & $p(z)$\\
F150DA-015 & 1.1064458e+02 & -7.3491646e+01 & 11.80 & WDF-P-4197 & $7.17_{-5.18}^{+4.08}$ & 0.53 & $p(z)$\\
F150DA-018 & 1.1069163e+02 & -7.3490501e+01 & 6.40 & WDF-P-4934 & $1.81_{-0.41}^{+0.43}$ & 0.00 & $p(z)$\\
F150DA-019 & 1.1062254e+02 & -7.3489227e+01 & 11.60 & WDF-P-2702 & $1.27_{-0.18}^{+0.20}$ & 0.00 & $p(z)$\\
F150DA-020 & 1.1069055e+02 & -7.3488899e+01 & 11.20 & WDF-P-5172 & $4.15_{-3.54}^{+0.47}$ & 0.00 & $p(z)$\\
F150DA-026 & 1.1064958e+02 & -7.3486969e+01 & 11.00 & WDF-P-3551 & $0.41_{-0.16}^{+0.20}$ & 0.00 & $p(z)$\\
F150DA-027 & 1.1071288e+02 & -7.3486656e+01 & 7.40 & WDF-P-5532 & $0.67_{-0.37}^{+0.25}$ & 0.00 & $p(z)$\\
F150DA-031 & 1.1062753e+02 & -7.3484474e+01 & 11.40 & WDF-P-2555 & $3.36_{-0.96}^{+8.78}$ & 0.41 & $p(z)$\\
F150DA-036 & 1.1071104e+02 & -7.3481148e+01 & 11.00 & WDF-P-4620 & $2.88_{-2.32}^{+1.35}$ & 0.03 & $p(z)$\\
F150DA-038 & 1.1072043e+02 & -7.3480797e+01 & 13.40 & WDF-P-4785 & $0.71_{-0.27}^{+0.52}$ & 0.00 & $p(z)$\\
F150DA-039 & 1.1071104e+02 & -7.3481148e+01 & 7.40 & WDF-P-4620 & $2.88_{-2.32}^{+1.35}$ & 0.03 & $p(z)$\\
F150DA-047 & 1.1066683e+02 & -7.3481430e+01 & 7.40 & WDF-P-3289 & $8.00_{-5.54}^{+7.32}$ & 0.56 & $p(z)$\\
F150DA-050 & 1.1064584e+02 & -7.3478386e+01 & 13.40 & WDF-P-2190 & $4.60_{-3.51}^{+5.22}$ & 0.34 & $p(z)$\\
F150DA-054 & 1.1062034e+02 & -7.3476067e+01 & 11.40 & WDF-P-930 & $7.93_{-5.55}^{+8.37}$ & 0.57 & $p(z)$\\
F150DA-057 & 1.1070329e+02 & -7.3475700e+01 & 11.40 & WDF-P-3537 & $3.84_{-3.24}^{+0.71}$ & 0.02 & $p(z)$\\
F150DA-058 & 1.1065908e+02 & -7.3475243e+01 & 13.40 & WDF-P-2102 & $2.98_{-2.61}^{+0.71}$ & 0.00 & $p(z)$\\
F150DA-060 & 1.1062808e+02 & -7.3473839e+01 & 11.40 & WDF-P-822 & $6.97_{-5.09}^{+4.58}$ & 0.53 & $p(z)$\\
F150DA-062 & 1.1068426e+02 & -7.3474068e+01 & 11.40 & WDF-P-2694 & $1.47_{-0.55}^{+0.75}$ & 0.00 & $p(z)$\\
F150DA-063 & 1.1068298e+02 & -7.3474297e+01 & 7.40 & WDF-P-2691 & $3.51_{-2.21}^{+6.45}$ & 0.20 & $p(z)$\\
F150DA-066 & 1.1062359e+02 & -7.3470810e+01 & 11.40 & WDF-P-178 & $13.82_{-9.12}^{+9.88}$ & 0.77 & $p(z)$\\
F150DA-077 & 1.1067607e+02 & -7.3466461e+01 & 13.40 & WDF-P-1210 & $15.00_{-10.20}^{+10.19}$ & 0.78 & $p(z)$\\
F150DA-078 & 1.1066418e+02 & -7.3464508e+01 & 11.80 & WDF-P-430 & $2.79_{-1.51}^{+2.46}$ & 0.00 & $p(z)$\\
F150DA-081 & 1.1066294e+02 & -7.3463394e+01 & 13.40 & WDF-P-237 & $1.67_{-1.60}^{+0.52}$ & 0.00 & $p(z)$\\
F150DA-082 & 1.1067789e+02 & -7.3463089e+01 & 7.60 & WDF-P-658 & $8.38_{-6.63}^{+9.25}$ & 0.58 & $p(z)$\\
F200DA-033 & 1.1064064e+02 & -7.3488998e+01 & 6.40 & WDF-P-3653 & $7.33_{-5.24}^{+6.28}$ & 0.57 & $p(z)$\\
F200DA-034 & 1.1073029e+02 & -7.3488083e+01 & 19.80 & WDF-P-6862 & $16.58_{-11.24}^{+7.92}$ & 0.80 & $S/N_{\rm det}=2.5<4$\\
F200DA-040 & 1.1072467e+02 & -7.3485855e+01 & 20.00 & WDF-P-6327 & $7.47_{-5.45}^{+6.90}$ & 0.54 & $p(z)$\\
F200DA-056 & 1.1061219e+02 & -7.3479179e+01 & 15.60 & WDF-P-1195 & $2.91_{-1.90}^{+6.64}$ & 0.19 & $p(z)$\\
F277DA-001 & 1.1065731e+02 & -7.3502014e+01 & -99.00 & WDF-P-7484 & $0.32_{-0.18}^{+0.16}$ & 0.00 & $p(z)$\\
F277DA-033 & 1.1070955e+02 & -7.3476181e+01 & -99.00 & WDF-P-3715 & $1.55_{-0.34}^{+0.80}$ & 0.00 & $p(z)$\\
F277DA-040 & 1.1060577e+02 & -7.3473579e+01 & -99.00 & WDF-P-50 & $0.66_{-0.25}^{+0.26}$ & 0.00 & $p(z)$\\
F277DA-044 & 1.1069387e+02 & -7.3470459e+01 & -99.00 & WDF-P-2421 & $0.09_{-0.06}^{+0.22}$ & 0.00 & $p(z)$\\
F277DA-045 & 1.1067736e+02 & -7.3469704e+01 & -99.00 & WDF-P-1729 & $3.81_{-1.01}^{+1.34}$ & 0.01 & $p(z)$\\
\enddata
\tablecomments{
}\label{tab:liter3}
\end{deluxetable*}

%%%%%%%%%%%%%%%%%%%%
\bibliography{sample631}{}
\bibliographystyle{aasjournal}

\end{document}